\documentclass[12pt,aps,twocolumn,pra, superscriptaddress,reprint,floatfix]{revtex4-1}
\usepackage{graphicx}
\usepackage{setspace}
\usepackage{color}
\usepackage{tabularx}
\usepackage{multirow}
\usepackage{tabularx}
\usepackage{float}
\usepackage{rotating}

\usepackage{booktabs}
\usepackage{lipsum}
\usepackage{siunitx}
\usepackage{bm}
\usepackage{mwe}
\usepackage{amsmath,amssymb}
\newsavebox{\measurebox}

\newcommand{\bea}{\begin{eqnarray}}
	\newcommand{\eea}{\end{eqnarray}}
\newcommand{\bes}{\begin{subequations}}
	\newcommand{\ees}{\end{subequations}}
\usepackage{stackengine}

\usepackage{BOONDOX-cal}
\usepackage{rotating}
\usepackage{setspace}
\usepackage{longtable}
\begin{document}
\bibliographystyle{revtex4-1}
\title{Modulational instability in $\cal{PT}$-symmetric  Bragg grating structures with saturable nonlinearity}
\author{K. Tamilselvan$^\dagger$}%
\affiliation{Department of Nonlinear Dynamics, School of Physics, Bharathidasan University, Tiruchirappalli - 620 024, India}
\author{A. Govindarajan$^\dagger$}
\email[Corresponding author: ]{govin.nld@gmail.com\\$\dagger$ Both authors contributed equally}
\affiliation{Department of Nonlinear Dynamics, School of Physics, Bharathidasan University, Tiruchirappalli - 620 024, India}
\author{I. Inbavalli}
\affiliation{Department of Physics, Presidency College (Autonomous), Chennai - 600 005, India}
\author{T. Alagesan}
\affiliation{Department of Physics, Presidency College (Autonomous), Chennai - 600 005, India}
\author{M. Lakshmanan}
\affiliation{Department of Nonlinear Dynamics, School of Physics, Bharathidasan University, Tiruchirappalli - 620 024, India}
\begin{abstract}
We investigate the nontrivial characteristics of modulational instability (MI) in a system of Bragg gratings with saturable nonlinearity. We also introduce an equal amount of gain and loss into the existing system which gives rise to an additional degree of freedom, thanks to the concept of  $\cal PT$- symmetry. We obtain the nonlinear dispersion relation of the saturable model and discover that such dispersion relations for both the conventional and $\cal PT$- symmetric cases contradict with the conventional Kerr and saturable systems by not displaying the typical signature of loop formation in either the upper branch or lower branch of the curve drawn against the wavenumber and detuning parameter. We then employ a standard linear stability analysis in order to study the MI dynamics of the continuous waves perturbed by an infinitesimal perturbation. The main objective of this paper is twofold. We first investigate the dynamics of the MI gain spectrum at the top and bottom of the photonic bandgap followed by a comprehensive analysis carried out in the anomalous and normal dispersion regimes. As a result, this perturbed system driven by the saturable nonlinearity and gain/loss yields a variety of instability spectra, which include the conventional side bands, monotonically increasing gain, the emergence of a single spectrum in either of the Stokes wavenumber region, and so on. In particular, we observe a remarkably peculiar spectrum, which is caused predominantly by the system parameter though the perturbation wavenumber boosts the former. We also address the impact of all the physical parameters considered in the proposed model which include coupling coefficient, dispersion parameter, and saturable nonlinearity on the phenomenon of MI for different $\cal PT$- symmetric regimes ranging from unbroken to broken one in greater detail. 

\end{abstract}
\maketitle
\section{Introduction}
Two decades ago, Bender and Boettcher introduced the concept of parity ($\cal P$) and time ($\cal T$) -symmetry to prove the conjecture which was put forward by Bessis in field theory (see \cite{Ben1998, Ben2002,Ben2007} and references therein), which then stimulated unprecedented interest across a wide range of fields in physics, including optics \cite{Gan2007, Makris2008, Guo2009,Makris2010, Rut2010,govindarajan2018}, photonics \cite{Reg2012,	Weimann}, and condensed matter physics \cite{Li2019}. In quantum mechanics, the concept was introduced to demonstrate that a non-Hermitian Hamiltonian can admit eigenvalues of real energy spectra when the operators $\cal P$  and $\cal T$ are performed simultaneously. It was later demonstrated that the realization of the $\cal PT$-symmetry  can be achieved in optics by a practical inclusion of a complex refractive index with equal gain /loss profile in optical systems {\color{blue}\cite{Gan2007}}. As is well-known, the distribution of complex refractive index in an optical system is defined by $n(x)=n_R(x)+in_I(x)$,  meaning that the index profile will be $\cal PT$-symmetric if the real and imaginary components of the index profile are respectively equal to an even function $n_{R}(x)=n_{R}(-x)$ and an odd function $n_{I}(x)=-n_{I}(-x)$ \cite{Gan2007, Makris2008, Rut2010}. This concept has been experimentally observed in several physical settings, including coupled waveguides and synthetic photonic lattices \cite{Rut2010,Reg2012}. In non-Hermitian $\cal PT$-symmetry, there exists a distinct phase-transition point at which the eigenvalue shifts from real to imaginary in its energy levels. \cite{Rut2010}. In the presence of a certain $\cal PT$-symmetric threshold level, in particular, below the phase-transition point, the system becomes stable, referred to as the unbroken $\cal PT$-symmetric regime, and above the $\cal PT$-threshold level, the system exhibits an exponential growth in its energy, thereby leading the former to an unstable state and eventually leading to spontaneous symmetry breaking. This unstable region is then known as the broken $\cal PT$-symmetric regime \cite{Pha2015}. The phase transition of $\cal PT$-symmetry is the key phenomenon underpinning the existence of different kinds of unusual dynamics, including non-reciprocity \cite{Makris2010} and double refraction \cite{Makris2008}. These advances in $\cal PT$-symmetry would enable the design of novel artificial optical systems that include periodic optical systems involving optical gain, and loss profile, periodic lattices, coupled structures, and passive experimental arrangements \cite{Konotop2016, Kotto2010, Suchkov}. 

The propagation of light in periodic structures, particularly in fiber Bragg gratings (FBGs), offers a number of highly versatile platforms for achieving a wide range of lightwave telecommunication applications. These include wavelength-stabilized pump lasers, dispersion compensators, narrow-band filters, and add-drop multiplexers \cite{Kas2010, Giles} to name a few. In the conventional FBG, the refractive index of the fiber is systematically altered by an intra-core Bragg grating, while the FBG operates in the ultraviolet region \cite{Hil1997, Kas2010}. To date, two distinct theories have been proposed to explain the propagation of light in FBGs: standard coupled-mode theory, which explains the propagation of forward and backward waves in a variety of distributed feedback structures \cite{Russell} and less-known Bloch-wave theory, that describes electron motion in semiconductors \cite{Hans}. The photonic bandgap, as in the linear domain, refers to the amount of chromatic dispersion present in the FBG, also known as the \textit{stopband} since the light transmission over a range of  frequency is restricted by the bandgap region \cite{Hans, Yariv, Kivshar}. There are many exciting phenomena that result from the combination of linear dispersion resulting from the band gap and nonlinearity introduced by the waveguide material, including Bragg solitons \cite{soliton}, gap solitons \cite{Taverner,Vignesh1}, optical switching \cite{switching}, pulse compression \cite{combression}, optical bi- and multi-stability \cite{bistability}, and modulational instability \cite{MI}. Recently, the notion of $\cal PT$-symmetry has been realized in FBGs in order to demonstrate the dynamic behavior of localized structures \cite{Miri}. The  formation and stable dynamics of solitons in $\cal PT$-symmetric FBGs  for different kinds of  linear effects and corresponding spectra have been extensively studied \cite{Suchkov}. Also, exactly solvable Dirac Hamiltonians are constructed by employing confluent Crum-Darboux transformation in the $\mathcal{PT}$- symmetric Bragg gratings \cite{FC}. Moreover, the concept of $\cal PT$- symmetry provides a stable platform for the development of various intriguing features, including unidirectional wave transport at exceptional points \cite{Lin2011}, coherent complete absorption in coupled resonators \cite{sun2014}, and nonreciprocal dynamics \cite{Makris2010} and so on. The role of $\cal PT$-symmetry has also been realized in linear FBG settings as a result of periodic modulation of the index and gain profiles which lead to a number of exotic dynamics such as asymmetrical mode coupling manifesting in unique reflection and transmission spectra. Particularly, there is a single reflection peak if the light is launched from the left end, while it becomes transparent on the rear end, which means that it does not reflect at all \cite{Poladian,Kulishov2005}. 

Modulation instability (MI) is a phenomenon that precedes the formation of localized modes in almost all nonlinear media, including FBGs. It is well-known that MI leads to an exponential growth of continuous wave (CW) as a consequence of a small perturbation imposed on it, and eventually, it breaks up into a train of localized ultra-short pulses \cite{Tai, Hasegawa}. This phenomenon originated in fluid dynamics as Benjamin-Feir instability \cite{Ben1967}, and then spread to a  variety of other fields, including optics \cite{Tai}, solid-state physics \cite{Remoi99}, plasma physics \cite{Akhtar2017}, and electrical lines \cite{Remoi99}. As a result of MI dynamics, prominent light-matter interactions are stimulated, which include phenomena such as Fermi-Pasta-Ulam-Tsingu recurrence in optics \cite{Simaeys}, and formation of Akhmediev breathers and Peregrine solitons \cite{Erkintalo} which can further be witnessed through the nonlinear stage of MI in optical fibers \cite{Kraych,Zakharov}. The implementation of MI could be utilized to achieve several potential applications, including the generation of ultrashort pulse trains at terahertz frequencies with high repetition rates \cite{Greer}, the generation of supercontinuum \cite{Dudley}, and the development of optical frequency combs \cite{LeoFC,Del}. These intriguing characteristics of MI phenomenon in fiber optics have been investigated both theoretically and analytically in a wide range of nonlinear media and have also been extended to emerging areas such as negative index materials, and $\cal PT$-symmetric media \cite{ZakharovMI}.

It is important to note that the studies on MI have been quite extensively investigated in the conventional FBGs, such as Kerr and non-Kerr nonlinear media, and apodized grating structures in which the CW states are then converted into a train of ultra-short pulses \cite{MI, Litchinitser2001,Ancima2009}. As with other nonlinear media, saturable nonlinear media play a prominent role not only in the formation of stable solitons but also in the MI dynamics,  where the nonlinear saturable parameter has a substantial impact on the MI gain spectrum, and its bandwidth. By and large, the refractive index of a saturable nonlinear medium increases with the intensity $I$, whereas it becomes saturated when the system is exposed to a sufficiently high level of input intensity \cite{Louis, Hick}.  In a Kerr-like medium, this saturable nonlinearity is typically characterized by its nonlinear refractive index ($n_{nl}(I)$) profile, such as $n_{nl}(I)=n_{\infty}(1-1/(1+I/I_{sat}))$ (where $I_{sat}$ and $n_{\infty}$ indicate the saturation intensity and the maximum change in the refractive index, respectively). A considerable amount of attention has been paid to this type of nonlinearity \cite{Soto,Higher}. Particularly, the saturable nonlinearity plays a very important role in preventing the catastrophic collapse of the nonlinear Schr\"odinger equation in higher dimensions \cite{Higher}. Also, such a saturable nonlinear medium supports the existence of stable localized solitons in various physical settings, such as two-level atomic systems \cite{Coutaz} and photorefractive materials like photovoltaic $LiNbO_3$ \cite{chenpr}. Along these lines, the impact of  saturable nonlinearity has been thoroughly analyzed on the ubiquitous process of MI  in the framework of nonlinear optics, in particular, semi-conductor-doped glass fibers and optical fibers  \cite{Hick, Silva, Dinda}. In the context of FBGs, the saturable nonlinearity has also been used to study the existence and stability of various types of solitons. The physical mechanism to realize such a model in the spatial region has also been put-forward by using a planar waveguide composed of photorefractive material with a longitudinal diffraction lattice written in its cladding \cite{Model}.
Although there exists a bunch of research works on the conventional FBGs, barring a single work \cite {sarma}, there seems to be no research work dealing with the study of MI in the $\mathcal{PT}$- symmetric FBG  in the literature. In addition, there is no study emphasizing the impact of saturable nonlinearity on the MI dynamics both in the conventional and $\mathcal{PT}$- symmetric media. Given these considerations in mind, in this paper, we carry out an extensive study elucidating the importance of saturable parameters both in the conventional and $\mathcal{PT}$- symmetric settings.

Following the detailed explanation of the proposed model in Sec. II, the analytical procedure of the linear stability analysis which is employed for the investigation of MI is provided in Sec. III. In Sec. IV, we then examine the dynamics of MI gain spectrum near the edges of the photonic bandgap and normal and  anomalous dispersion regimes. Each of these cases is analyzed for the effect of various physical parameters, including gain/loss and saturable nonlinearity under different $\cal PT$- symmetric regimes. We conclude our findings with a detailed summary in Sec. V.
\begin{figure*}[t]
	\centering\includegraphics[width=6cm,height=5cm]{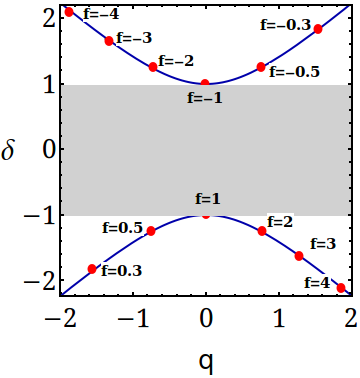} 
	\caption{(Color online) The dispersion curve plotted between $\delta$ and  $\mbox{q}$  for linear grating structure. Here $f=\mp1$ refers to the top and bottom of the photonic bandgap structure with the system parameter, $\kappa=1$ and
$g=0$.} \label{Fig1}
\end{figure*}

\section{Model}
We consider $\cal PT$-symmetric fiber Bragg gratings with a period $\Lambda$ imprinted on the core of the fiber of refractive index $n_{0}$ and length $z$. Mathematically, the distribution of the refractive index $(n(z))$ profile for a $\mathcal{PT}$-FBG with saturable nonlinear media can be described as follows \cite{Vraja}:
\bea\label{refractive}
&&n(z)=n_{0}+n_{1R} \cos \left(\frac{2 \pi z}{\Lambda}\right)+i n_{1I} \sin \left(\frac{2 \pi z}{\Lambda}\right)\nonumber\\ 
&&-n_{2} \mathcal{f}(|E|^2).
\eea
Here, $n_{1}$ represents the strength of the modulation parameter and its real and imaginary parts are, respectively, indicated through terms $n_{1R}$ and $n_{1I}$, which are responsible for the $\cal PT$-symmetric potential. The term $n_{2}$ represents the nonlinear refractive index pertaining to the saturable nonlinearity of the Bragg structure. The function $\mathcal{f}(|E|^2)$ can be expressed as $1/(1+|E|^{2})$, where $E$ stands for the optical field \cite{Model}. By taking the square on the above nonlinear refractive index profile \eqref{refractive} and neglecting the higher order terms in $n_{1}$ and $n_{2}$, one can obtain the following reduced form  
\bea\label{squarerefractive}
&&n^{2}(z)=n^{2}_{0}+2 n_{0} n_{1R} \cos \left(\frac{2 \pi z}{\Lambda}\right)+2in_{1R}n_{1I}\sin \left(\frac{2 \pi z}{\Lambda}\right)\nonumber\\ 
&&-2 n_{0} n_{2} \mathcal{f}(|E|^{2}).
\eea
The dynamics of the system under study can be modeled by the following time-dependent Helmholtz equation, for the optical field $E$ as.
\bea \label{Helmholtz}
\frac{\partial^{2} E}{\partial z^{2}}+\frac{\partial^{2} E}{\partial t^{2}}+k^{2}\frac{n^{2} (z)}{n_{0}^{2}}E=0,
\eea
where $k$ represents the wave vector. We seek an optical field ($E(z,t)$) that consists of forward and backward components propagating inside  the FBG as follows:
\bea \label{Seedsolution}
E(z,t)=\Psi_{1}(z,t) \exp(i (k z-\omega_{0} t)+\nonumber\\ \Psi_{2}(z,t) \exp(-i (k z-\omega_{0} t)),
\eea
where the terms $\Psi_{1}(z,t)$ and $\Psi_{2}(z,t)$ indicate slowly varying amplitudes of the forward and backward electric fields, respectively and $\omega_{0}$ indicates the frequency of the incident light. Substituting Eq. \eqref{Seedsolution} into Eq. \eqref{Helmholtz} and applying the synchronous approximation, one can obtain the normalized  coupled-mode equations with the saturable nonlinearity and equal amounts of gain and loss as follows \cite{Model,Vignesh1},  
\bes\label{PTCoupler}\bea
i \left(\frac{\partial \Psi_{1}}{\partial z}+ \frac{1}{v}  \frac{\partial \Psi_{1}}{\partial t}\right)+(\kappa +g)\Psi_2-\mathcal{f}(|\Psi_{1,2}|) \Psi_1=0,\quad\\ \label{1a}
-i \left(\frac{\partial \Psi_{2}}{\partial z}-\frac{1}{v} \frac{\partial \Psi_{2}}{\partial t}\right)+(\kappa-g) \Psi_1-\mathcal{f}(|\Psi_{1,2}|) \Psi_2=0.\quad\label{1b}
\eea\ees
where the group velocity of light is given by $v_g$=$(c/n)$. Here $c$ indicates the speed of light, and the nonlinear term $\mathcal{f}(|\Psi_{1,2}|^2)$ can be expressed as  $\Gamma/(1+|\Psi_{1}|^2+|\Psi_{2}|^2)$, in which $\Gamma=2\pi n_2/\lambda_{0}$ refers to the strength of the saturation parameter, where $\lambda_{0}$ refers to the wavelength in free space. For a detailed derivation of the theoretical model, one may refer to Refs. \cite{Model,Vignesh1}. Following the general settings, the total intensity of light can be calculated using the expression $I=\sum_{j=1}^{2}|\Psi_{j}|^{2}$. In Eq.~\eqref{PTCoupler}, $\kappa=\pi n_{1 R} / \lambda_{0}$ and $g=\pi n_{1 I} /\lambda_0$, respectively, attribute to the linear coupling coefficient and gain and loss profile of the fiber Bragg grating system. Based on the parameters $g$ and $\kappa$, three distinct $\cal PT$- symmetric conditions can be formulated. For instance, when $g=\kappa$, we obtain a unique exceptional point, (also known as the $\cal PT$-symmetric threshold level), while $g<\kappa$ leads to the broken $\cal PT$-symmetric regime. Similarly, the condition $g>\kappa$ is known as the unbroken $\cal PT$- symmetric regime as the system tends to show stable dynamics in this case \cite{Liu2015}. It is to be noted that these classifications are utilized to describe the characteristics behavior of MI in four different domains, including the conventional case ($g=0$), below, at, and above the $\cal PT$-symmetric regimes for Eq. \eqref{PTCoupler}, in the present study. 
\begin{figure*}[t]
\topinset{(a)}{\includegraphics[width=5.5cm,height=5.2cm]{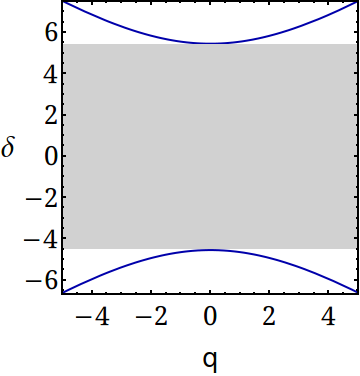}}{0.1in}{-0.43in}~
\topinset{(b)}{\includegraphics[width=5.5cm,height=5.2cm]{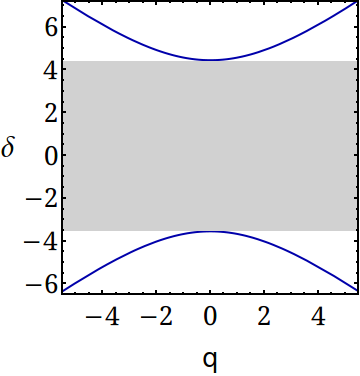}}{0.1in}{-0.43in}
\topinset{(c)}{\includegraphics[width=5.3cm,height=5.35cm]{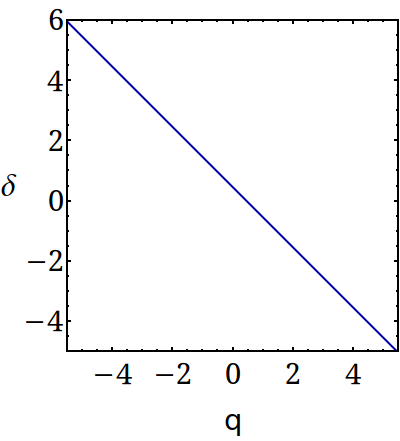}}{0.1in}{-0.45in}
\topinset{(d)}{\includegraphics[width=5.5cm,height=5.3cm]{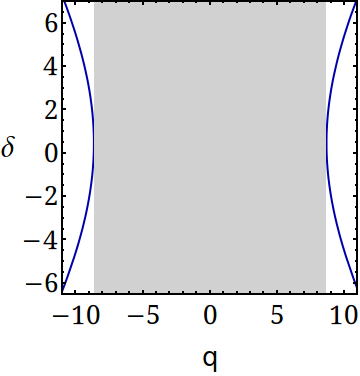}}{0.1in}{-0.45in}
	\caption{(Color online) The characteristics of nonlinear dispersion relation drawn between $\delta$ and $\mbox{q}$ for (a) Conventional $g=0$, (b) below $\cal PT$-symmetric threshold $g=3$ (c) at $\cal PT$ $g=5$, and (d) above $\cal PT$-symmetric threshold $g=10$. The rest of the parameters are $P=10$ and $\kappa=\Gamma=5$. }\label{Fig2}
\end{figure*}

Prior to performing a linear stability analysis of the proposed system \eqref{PTCoupler}, we wish to study the dispersion relations for both the conventional and $\cal PT$- symmetric cases. To this end, we consider the following counter-propagating continuous wave (CW) solutions: 
\bes\label{CW}\bea
\Psi_{1}= \alpha \exp(i (\mbox{q} z - \delta v t)), \\
\Psi_{2}= \beta \exp(i (\mbox{q} z - \delta v t)).
\eea\ees 
Here the parameters $\alpha$ and $\beta$ indicate the forward and backward wave amplitudes that are assumed to be real constants, and the total power of the grating structure is defined as $\alpha^{2}+\beta^{2}=P$, which can also be calculated from $|\Psi_{1}|^{2}+|\Psi_{2}|^{2}$. We now introduce a term that represents a ratio between the two constants as  $f=\beta/\alpha$, where  $\alpha=\sqrt{P/(1+f^2)}$ and $\beta=\sqrt{P~f^2/(1+f^2)}$. Substituting Eqs.~\eqref{CW} into the system \eqref{PTCoupler}, one can obtain a mathematical expression for nonlinear dispersion relations as given below.
\bes\label{nonlinear0}\bea
\delta&=&\frac{\Gamma}{(1+P)}-\frac{1}{2} \left(\frac{(\kappa+g)f^{2}+(\kappa-g)}{f}\right),\\
\mbox{q}&=&\frac{(\kappa+g)f^{2}-\kappa+g}{2f}.
\eea\ees
Before studying the nonlinear dispersion relation, it is always instructive to first look into the characteristics of the linear dispersion relation, which can be obtained in the following way by turning off the saturable nonlinear parameter ($\Gamma=0$) in Eq.~\eqref{PTCoupler}, so that from Eq. \eqref{nonlinear0}, we have 
\bes\bea\label{linear}
\delta&=&-\frac{1}{2} \left(\frac{(\kappa+g)f^{2}+(\kappa-g)}{f}\right),\\
\mbox{q}&=&\frac{(\kappa+g)f^{2}-\kappa+g}{2f}.
\eea\ees
By using Eqs. (8a) and (8b), the exact linear dispersion relation $\delta(\mbox{q})$ can be found as:
\bea\label{linear}
\mbox{q}^{2}&=&\delta^{2}-\kappa^{2}+g^{2}.
\eea
Similarly the exact nonlinear dispersion relation can be deduced as follows,
\bea\label{nonlinear}
\mbox{q}^{2}&=&\left(\delta-\frac{\Gamma}{1+P}\right)^{2}-\kappa^{2}+g^{2}.
\eea
Figure~\ref{Fig1}~reveals the photonic bandgap structure as a result of the linear dispersion relation of the Bragg grating. As there exists a photonic bandgap formed within a range of frequencies (which is also called forbidden frequencies), the propagation of light is restricted inside it, viz. between the upper and lower branches, and most of the light is reflected. Outside this region, light is allowed to traverse. Note that in a uniform medium without the presence of Bragg gratings, light propagates at its own speed. However, the inclusion of the Bragg grating exhibits a slowly decreasing dispersion at frequencies near the edges of the photonic bandgap, where  light experiences a slow propagation when compared to the uniform medium. 

The parameter $f$ is closely related to the group velocity in such a way that $v_g=d\delta/dq=[\kappa(1-f^2)-g(1+f^{2})]/[\kappa(1+f^2)-g(1-f^{2})]$. In particular, for $f=\pm1$, the edges of the upper and lower branches of the dispersion curve are very close together and $f=-1$ and $f=1$ are specified as the top and bottom of the photonic bandgap, respectively. Also, when $f<0$ it corresponds to the upper dispersion curve, the group velocity dispersion becomes negative indicating the anomalous dispersion regime. On the other hand, $f>0$ refers to the normal dispersion regime by the lower branch dispersion curve. 

Based on Eq.~\eqref{nonlinear}, we present the characteristics of the nonlinear dispersion in Fig.~\ref{Fig2} for the Bragg gratings under three different $\cal PT$-symmetric conditions with $P=10$ and $\kappa=\Gamma=1$. When $g=0$, which refers to the conventional case, there is a typical propagation of forward and backward wave vectors, as shown in Fig.~\ref{Fig2}(a), with a broad bandgap in which the formation of gap solitons would exist in the considered grating system. Upon increasing the value of $g$ further ($g=0.5$), i.e., below $\cal PT$-symmetric threshold, one can observe that the size of the bandgap has reduced considerably as compared to the previous conventional case, as seen in Fig.~\ref{Fig2}(b). In contrast to the prior cases, no dispersion curve and bandgap exist at the exceptional point ($\cal PT$ threshold, $g=1$), as witnessed in Fig.~\ref{Fig2}(c). Also, in this case, it is apparent that no localized structure, including gap solitons, can be observed.  The role of the dispersion curve in the broken $\cal PT$-symmetric regime is addressed in Figs.~\ref{Fig2}(d), where it is clear to observe that the characteristics of the dispersion relation are similar to those of the regions such as  conventional and  below the $\cal PT$-symmetric threshold, except that the dynamics of the forward and backward wave vectors have been shifted by $90^{\circ}$ clockwise. It is worth noting that the results corresponding to the nonlinear dispersion relations can be utilized to find Bragg and gap solitons in the nonlinear $\cal PT$-symmetric fiber Bragg gratings. Generally, the formation of a loop structure on either the upper or lower branches of a nonlinear dispersion curve is noticed in various types of periodic structures \cite{Litchinitser2001,Ancima2009, SNL1,zhong}. Nevertheless, the saturable $\cal PT$-symmetric FBG system under study never allows the formation of such a loop structure on the nonlinear dispersion curve due to the unique form of saturable nonlinearity considered here. In particular, the level of the nonlinearities (self- phase and cross-phase modulations) has been assigned with a ratio of 1:1 in the present system (5) as opposed to the 1:2  ratio which is generally adopted in the Kerr-like Bragg grating structures \cite{Litchinitser2001,Ancima2009, SNL1,zhong}. As a result, this new-ratio has changed the nonlinear dispersion relation in a way that the nonlinearity parameter $\Gamma$ is present in Eq.~\ref{nonlinear}(a) alone instead of appearing in both the dispersion relations, which in turn as one of the reasons for not supporting the formation of the loop in the dispersion curves.

For the investigation of the characteristic behavior of MI for the proposed system \eqref{PTCoupler}, we employ the standard method, namely the linear stability analysis (LSA) in the next section.

\section{Linear Stability Analysis}
In the linear stability analysis,  infinitesimal perturbations are imposed on the CW state that result in the exponential growth of its amplitude. Based on the general procedure, let us consider the CW solutions \eqref{CW} with small perturbations as follows,
\bes\label{CW1}\bea
\Psi_{1}= \left(\alpha+a_{1}(z,t)\right) \exp(i (\mbox{q} z - \delta v t)), \\
\Psi_{2}= \left(\beta +a_{2}(z,t)\right)\exp(i (\mbox{q} z - \delta v t)).
\eea\ees 
The functions $|a_{1,2}|$ ($\ll \alpha, \beta$) are the perturbations imposed on the steady-state solution. Substituting Eq. \eqref{CW1} into Eq. \eqref{PTCoupler} and linearizing with respect to  the perturbations $a_{1,2}$, one can obtain the following equations. 
\bes\label{linearPTCoupler}\bea
&&i a_{1,z}+ \frac{i}{v} a_{1,t}+ \epsilon_{1} a +\epsilon_{2} a_2+\epsilon \left( a_{1}^{*}+f a_2^{*}\right)=0,\qquad\\
&&-i a_{2,z}+\frac{i}{v} a_{2,t}+ \epsilon_{3}a_2+\epsilon_{4} a_1+ \epsilon \left(fa_1^{*}+f^{2}a_{2}^{*}\right)=0,\qquad\quad
\eea\ees
where
\bea 
&&\epsilon=\frac{\Gamma P}{(1+f^2)},\,\, \epsilon_{1}=\epsilon-f(\kappa +g),\,\,\epsilon_{2}=\kappa+g+f \epsilon,\nonumber\\
&&\epsilon_3= \frac{g-\kappa}{f}+f^{2}\epsilon ,\,\,\epsilon_{4}=f\epsilon+(\kappa -g).                
\eea
Next, we consider the Fourier components of the perturbed CW amplitudes $a_{1,2}(z,t)$ as

\bes\bea\label{PertEq1}
&&a_{1}(z,t)=\mbox{p}_{+} e^{i(K z-\Omega t)}+\mbox{p}_{-} e^{-i(K z-\Omega t)},\\
&&a_{2}(z,t)=\mbox{q}_{+} e^{i(K z-\Omega t)}+\mbox{q}_{-} e^{-i(K z-\Omega t)}.\label{PertEq2}
\eea\ees
Here, $\mbox{p}_{+}$ and $\mbox{q}_{+}$  represent the forward propagation, whereas $\mbox{p}_-$ and $\mbox{q}_-$ represent the backward propagation. Also, $K$ and $\Omega$ indicate the wavenumber and frequency of the perturbation, respectively.  In what follows, on the basis of the terminology adopted in the study of light scattering  in quantum mechanics \cite{Boyd}, we will refer to the region in which the wave number takes negative values $(K<0)$ as the Stokes wavenumber region and the other region in which the wave number is positive $(K>0)$ is known as the anti-Stoke wavenumber region \cite{Pan}. Substituting the above solutions \eqref{PertEq1} and \eqref{PertEq2} into the governing equations \eqref{PTCoupler} and linearizing the resultant equations with the perturbed amplitudes $\mbox{p}_{+}$ and $\mbox{p}_{-}$ and $\mbox{q}_{+}$ and $\mbox{q}_{-}$, one obtains four homogeneous equations in the following matrix form:
\bea\label{matrix}
[Y]\times[u]^{T}=0, \qquad\qquad u^{T}=(\mbox{p}_+,\mbox{q}_+,\mbox{p}_-,\mbox{q}_-)\
\eea
where $Y$ is a $4\times4$ matrix with the following elements,
\bea
&&y_{11}= -f (\kappa+g)-K +\epsilon +\Omega,y_{12}=\epsilon, y_{13}=\kappa+g+f \epsilon,\nonumber\\
&&y_{14}=f \epsilon,y_{21}=\epsilon,y_{22}=-f(\kappa+g)+K +\epsilon-\Omega,y_{23}=f \epsilon\nonumber\\
&&y_{24}=g +\kappa+f \epsilon,y_{33}=((g-\kappa)/f)+K + f^{2} \epsilon- +\Omega\nonumber\\&& y_{31}=\kappa-g +f \epsilon,y_{32}=f\epsilon,y_{34}=f^{2} \epsilon ,y_{41}=f \epsilon,y_{43}=f^{2} \epsilon,\nonumber\\
&&y_{42}=\kappa-g+f \epsilon,y_{44}=((g-\kappa)/f)-K + f^{2} \epsilon-\Omega.
\eea
It is important to note that the $Y$ matrix has non-trivial solutions when its determinant vanishes, which in turn leads to the following quartic polynomial equation in $\Omega$:
\bea
\Omega^{4}+a \Omega^{2} + b \Omega + c=0,
\eea
where
\bea
&&a=-\frac{K^{2}d_{2} g^2+2 e_{1} g \kappa+\kappa^2+f^4 \kappa^2+2f^2 e_3}{f^2},\nonumber\\
&& b=-\frac{2 K e_{1} g^{2}+ e_1 \kappa^2+2 g(-2 f^3 \epsilon+\kappa+f^4 \kappa)}{f^2},\nonumber\\
&&c=-\frac{ d_{1} g^2+2e_{1}g \kappa-2 f^3\epsilon \kappa+\kappa^2+f^4 \kappa^2-f^2 e_{2} }{f^2},\qquad\nonumber\\
&&d_{1,2}=(\pm 1+f^{2})^2,\, e_{1}=(-1+f^4),\,\nonumber\\
&&e_2=(K^{2}+2\kappa(3f \epsilon+\kappa)),\, e_3=(K^2+\kappa^{2}).
\eea
This quartic polynomial equation has four branches of the solution when $\Omega(K)$ satisfies the following relations,
\bea\label{wavenumber}
\Omega_{1-4}=\pm \frac{1}{2 \sqrt{6}}\left({\sqrt{sgn(\nu_{1})|\nu_{1}|}}\pm \sqrt{sgn(\nu_{2})| \nu_{2}|}\right),
\eea
where
\bea
&&\nu_{1}=-4a+\frac{2^{(1/3)}\Upsilon_{1}}{\Upsilon_{2}},\nonumber\\
&&\nu_{2}=-8a+\frac{\Upsilon_{1}}{\Upsilon_{2}}-2^{(1/3)}\Upsilon_{2}\pm\frac{12\times\sqrt{6}b}{\sqrt{\nu_{1}+8a+2^{(1/3)}\Upsilon_{2}}},\qquad\nonumber\\
&&\Upsilon_{1}=2\times2^{(1/3)}(a^{2} +12c),\nonumber\\
&&\Upsilon_{2}=(2a^3+27b^2-72 ac +\eta)^{(1/3)},\nonumber\\
&&\eta= \sqrt{-4(a^2 + 12 c)^3+(2 a^3 + 27 b^2-72ac)^2}.
\eea
It is to be noted that one can find the conditions for which the four branches of the above relation given in Eq. \eqref{wavenumber} become imaginary, which include,  
\begin{itemize}
\item (i) $sgn(\nu_{1})<0$ and  $sgn(\nu_{2})<0$, 
\item (ii) $sgn(\nu_{1})>0$ and  $sgn(\nu_{2})<0$, and 
\item (iii) $sgn(\nu_{1})<0$ and  $sgn(\nu_{2})>0$.
\end{itemize}
\begin{figure*}[t]
	\topinset{(a)}{\includegraphics[scale=0.42]{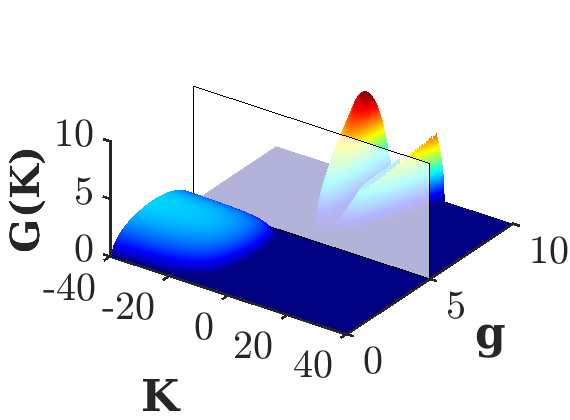}}{0.4in}{-0.45in}
	\topinset{\color{white}{(b)}}{\includegraphics[scale=0.37]{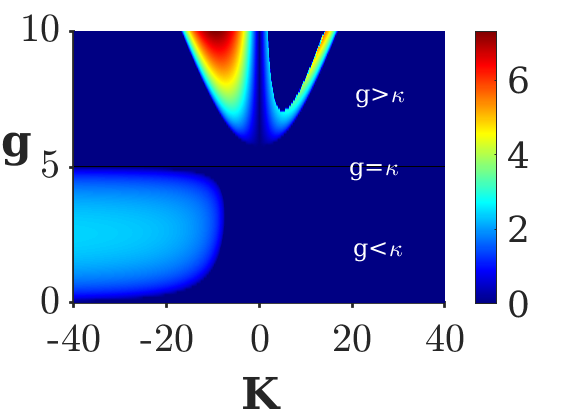}}{0.1in}{-0.55in}
		\topinset{(c)}{\includegraphics[scale=0.3]{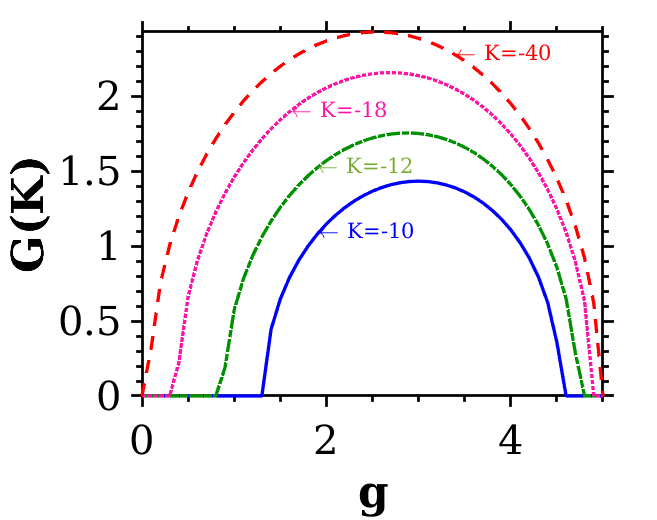}}{0.1in}{-0.4in}

	\caption{(Color online) (a) The MI gain spectra with the variation in $g$ at the bottom of the photonic bandgap $f=1$, (b) its corresponding contour view, and (c) The peak gain of the peculiar MI spectrum as a function of the gain/loss parameter for different values of wavenumber. The other parameters are  $\kappa=5$, $P=1$, and $\Gamma=5$. }\label{Figure3}
\end{figure*} 
Hence the growth rate of the MI gain spectrum $G(K)$ can be calculated using the relation $G(K)=Im(\Omega_{max})$, where $\Omega_{max}$ represents the largest imaginary part of the four branches. The purpose of this study is to examine the dynamics of CW instability in the two different dispersion regimes, namely the anomalous dispersion regime $(f<0)$ and the normal dispersion regime $(f>0)$. As a first step, we examine the MI gain spectrum emerging at the bottom and top of the photonic bandgaps. We will also analyze how the various system parameters, such as gain/loss, saturable nonlinear coefficient, and power, influence the MI spectrum in each of the $\cal PT$-symmetric domains.
\section{Investigations on Modulational Instability}
\subsection{Bottom of the photonic bandgap}
We here investigate the MI gain spectrum as a function of the gain and loss parameter $(g)$ at the bottom of the photonic bandgap ($f=1$).
When the value of the linear coupling coefficient is fixed at $\kappa=5$, the resultant MI gain spectrum is illustrated for the continuous variation of $g$ invoking all the three $\mathcal{PT}$- symmetric regimes in addition to the conventional case. The results are shown in Figs. \ref{Figure3}(a) and (b), where one can observe a peculiar spectrum for a certain range of $g$ ($0<g<4.9$) on only one side (left) of the zero value of perturbation wavenumber while on the other side, the system does not experience any instability at all by exhibiting no sideband there even after the addition of the perturbation with a wide range of input wavelengths. It is pertinent to note that the spectrum obtained in the unbroken $\cal PT$ -symmetric regime is unique in two aspects when compared to the standard spectrum obtained in conventional systems. First, although the growth rate of CW instability tends to rise gradually with the increase in the value of wavenumber, the exponential growth is predominantly noticed as a function of the gain/loss parameter, which is apparently seen in Fig. \ref{Figure3}(c). Second, the shape of the MI spectrum is different as the exponential growth is observed with the increase in the value of the gain/loss parameter with a peak gain in the middle of these values in contrast to the perturbation wavenumber. We would like to emphasize that the finding of such a unique spectrum is new and not reported in any conventional and $\cal PT$-symmetric FBGs as far as our knowledge goes. With a further increase in $g$ leading to the broken $\cal PT$- symmetric regime, it is quite interesting to observe the conventional MI gain spectrum (as obtained in the standard nonlinear Schr\"odinger equation in the \textit{anomalous dispersion regime}) appearing on either side of the perturbation wavenumber ($K$) for a certain (minimum) range of gain/loss parameter $g$ ($5.8<g<7$). Nevertheless, subsequently, it turns out to be a case of two distinct sidebands (asymmetric spectra) that include a huge gain spectrum with a wide bandwidth that appeared in the Stokes wavenumber region while a split-up occurs in the MI spectrum on the other side which eventually leads to two different additional spectra, including primary one found near the zero perturbation wavenumber with a lower gain and bandwidth and a secondary MI band which is far detuned with a comparatively higher gain and bandwidth.  
 \begin{figure}[t]
	\topinset{(a)}{\includegraphics[scale=0.25]{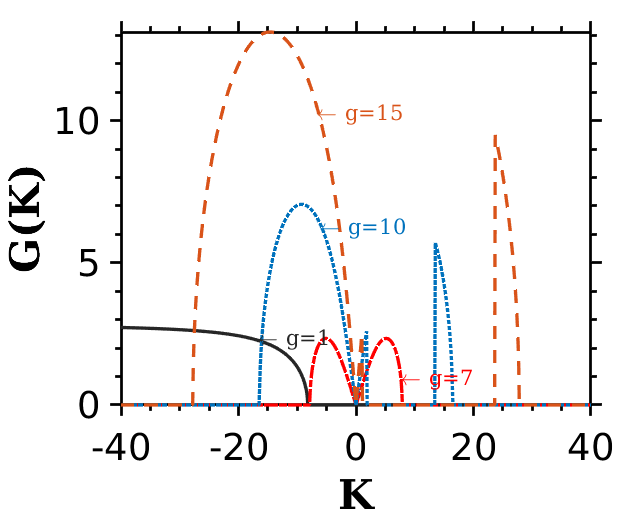}}{0.2in}{-0.4in}
	\topinset{(b)}{\includegraphics[scale=0.25]{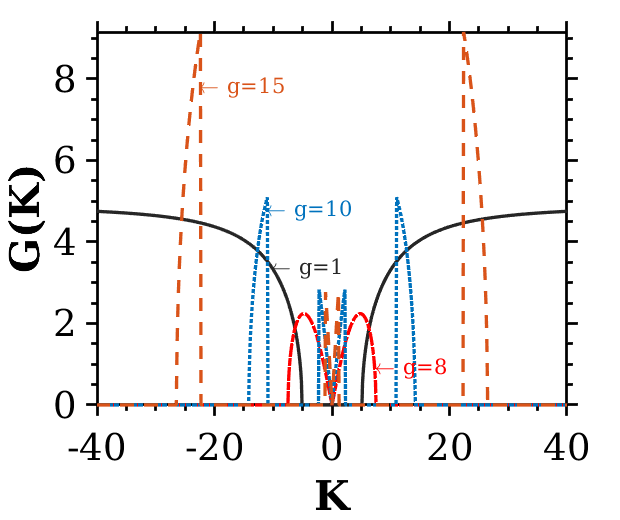}}{0.2in}{-0.4in}
	\caption{(Color online) One-dimensional MI spectra for different values of $g$ for (a) $\Gamma=6$ and (b) $\Gamma=10$. The parameters are assigned the values of $\kappa=5$, and $P=f=1$. }\label{Figure31} 
\end{figure} 

To gain further insight into the characteristic behavior of the instability spectra obtained at the bottom of the photonic bandgap, we present two additional plots in Fig. \ref{Figure31} that delineate the MI growth rates as a function of $g$ for two different values of the saturable nonlinearity, including $\Gamma=6$ and $\Gamma=10$. From Fig. \ref{Figure31}(a), it is clear that the structure of the MI spectrum changes differently as a result of tuning the parameter $g$ from the conventional case to different $\cal PT$- symmetric regimes. As pointed out earlier, here too, we notice the peculiar MI spectrum in the unbroken $\cal PT$-symmetric regime. Upon reaching the $\cal PT$-symmetric threshold condition, there are two different sidebands that got split up from the conventional one; the primary MI spectrum located further Stokes wavenumber regime and a secondary MI spectrum in the positive wavenumber region having a lower gain compared to the former. The gain and bandwidth of MI spectra are dramatically enhanced by one-third as much as in the former case when we increase the value of $g$ further ($g=15$). Furthermore, when we set the value of the nonlinear saturable parameter at $\Gamma=10$ (see Fig. \ref{Figure31}(b)), we observe a slightly modified MI gain spectrum compared to the previous one for each value of $g$. For example, the peculiar spectrum appears in the anti-Stokes wavenumber region too for $g=1$. Similarly, the MI band observed in the Stokes wavenumber region is also split-up into two different bands as noticed in the previous case, as shown in Fig.~\ref{Figure31}(a). In addition, the values of gain and bandwidth are significantly suppressed at each $g$ value relative to the previous case. These ramifications clearly reveal the fact that the role of saturable nonlinearity is intense in altering the MI spectrum in the presence of gain/loss parameter.   
\subsubsection{The impact of coupling coefficient $\kappa$ }
\begin{figure}[t]
	\topinset{(a)}{\includegraphics[scale=0.3]{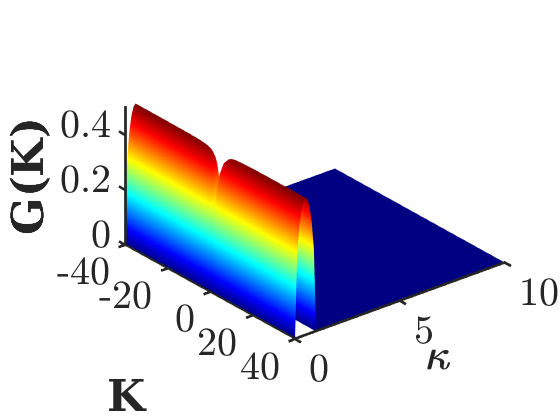}}{0.4in}{-0.4in}
 	\topinset{(b)}{\includegraphics[scale=0.23]{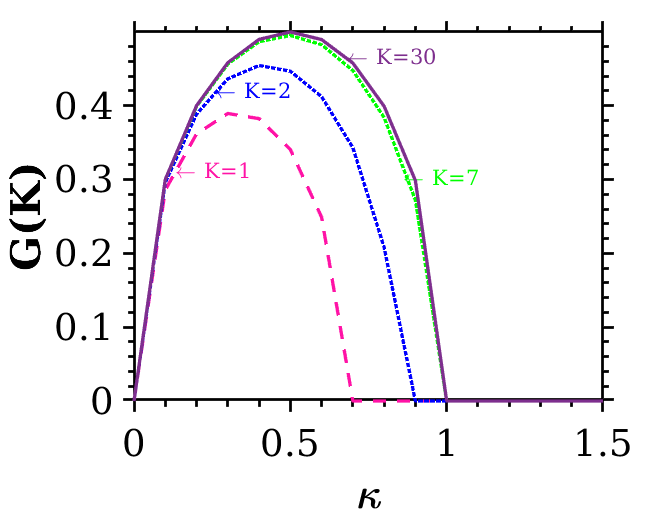}}{0.1in}{-0.38in}
	\caption{(Color online) (a) The MI gain spectra as a function of $\kappa$ at the bottom of the photonic bandgap $f=1$, for the conventional case ($g=0$). (b) The maximum gain of a peculiar MI spectrum versus the coupling coefficient $\kappa$ for distinct wavenumbers. Here the parameters are assigned as $P=1$, and $\Gamma=1$. }\label{NFigure1}
\end{figure} 
We now investigate the role of the coupling coefficient ($\kappa$) in the development of the MI gain spectrum at the bottom of the photonic bandgap ($f=1$) as shown in Fig. \ref{NFigure1}(a). Note that we here present only the conventional case by setting the gain/loss parameter to zero, since the different $\cal PT$-symmetric conditions exhibit the same MI pattern as witnessed in Fig.~\ref{Figure31} when we tune the value of gain/loss parameter. Now coming to the conventional system, the role of the coupling parameter results in the peculiar MI spectrum appearing on either side of the perturbation wavenumber ($K$) in the range $0<\kappa<1$.  By further increasing the value of $\kappa$, it is apparent that the CW exhibits complete stable propagation on both sides of the wavenumber without exhibiting any instability. 
Also,  in the light of this investigation, it can easily be interpreted that the CW propagation is unstable whenever the coupling coefficient ($\kappa$) has a non-trivial value (up to the unity in the normalized scale) in the conventional system for the given system parameters. It is also to be noted that the maximum MI gain of  the peculiar spectrum significantly increases as the wavenumber ($K$) increases, though the spectrum is obtained as a function of the coupling coefficient as seen in Fig.~\ref{NFigure1}(b). 

\subsubsection {Impact of power}
\begin{figure}[t]
	\topinset{(a)}{\includegraphics[scale=0.27]{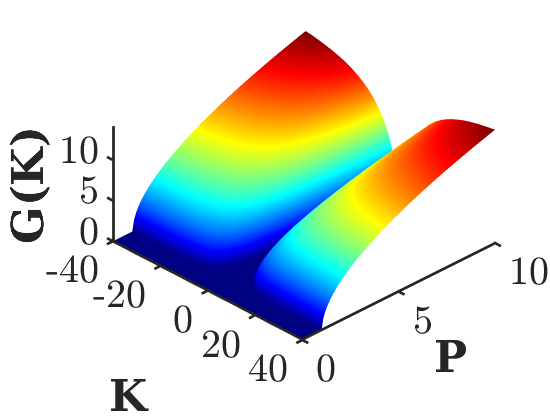}}{0.4in}{-0.4in}
	\topinset{(b)}{\includegraphics[scale=0.29]{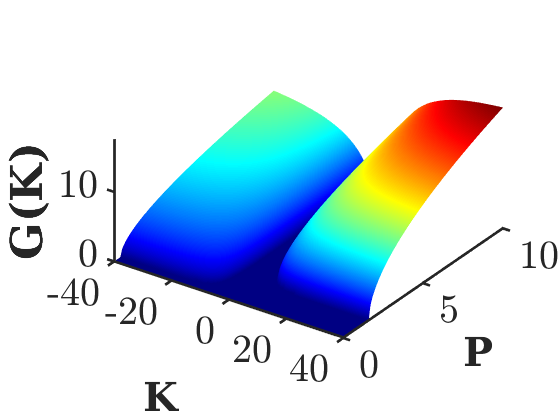}}{0.4in}{-0.4in}
	\topinset{(c)}{\includegraphics[scale=0.28]{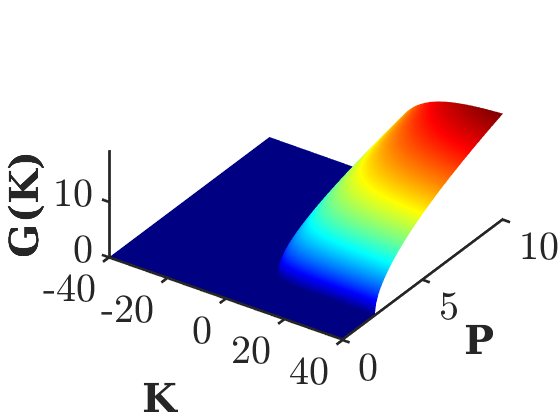}}{0.4in}{-0.4in}
	\topinset{(d)}{\includegraphics[scale=0.28]{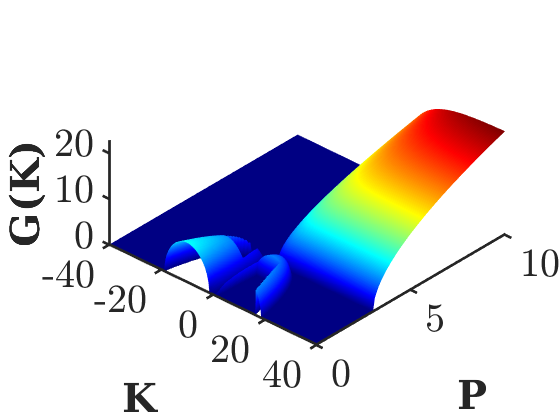}}{0.4in}{-0.3in}
	\caption{(Color online) The plots showing the MI gain spectra as a function of $P$ at the bottom of the photonic bandgap $f=1$ for four different cases. (a) conventional, (b) below, (c) at, (d) above $\cal PT$- symmetric thresholds. The parameters are assigned as $\kappa=5$, and $\Gamma=5$, and $f=1$. }\label{Figure4}
\end{figure}

This subsection examines how power affects the MI gain spectrum at the bottom of the photonic bandgap.  As can be seen in Fig.~\ref{Figure4}(a), in the conventional case $(g=0)$, a symmetric pattern of monotonically increasing side gain emerges on either side of the central perturbation wavenumber $(K=0)$, which appears to become wider with the increase in the value of $P$. There is also a significant separation distance between the monotonically increasing side gains. These symmetric MI patterns can be turned into asymmetric ones as shown in Fig.~\ref{Figure4}(b) by tuning the value of $g$ to $g=3$, i.e., below the $\cal PT$-symmetric threshold. Also, the increase in the value of $g$ results in a suppression of the monotonically increasing side-gain located in the Stokes wavenumber region, thereby leading to a drop in  its growth rate than that of the side-gain in the anti-Stokes wavenumber region. At the $\cal PT$- symmetric threshold regime (see Fig.~\ref{Figure4}(c)), monotonically increasing side-gain in the Stokes wavenumber region gets completely suppressed and the CW exhibits stable dynamics, whereas the side gain located in the anti-Stokes wavenumber region remained the same against the variation in $P$. Note that in all of the above situations, the side MI gain increases as the input power increases. In the case of broken $\cal PT$- symmetric regime, the MI spectrum qualitatively changes into an unusual one featuring two asymmetric MI sidebands on either side of the zero perturbation wavenumber as seen in Fig.~\ref{Figure4}(d). In particular, one can observe a huge MI gain spectrum with a wide bandwidth in the Stokes wavenumber region, and a set of two distinct spectra that include the primary MI sideband and secondary MI band in the anti-Stokes wavenumber region. With a small increase of $P$, these MI spectra tend to split in the side of Stokes wavenumber and try to merge together on the other side. Further increment in the value of $P$ combines all the former spectra into a single huge spectrum with a  monotonically increasing side gain in the anti-Stokes wavenumber region.
\subsubsection{MI as a function of saturable nonlinearity}
\begin{figure}[t]
	\topinset{(a)}{\includegraphics[scale=0.25]{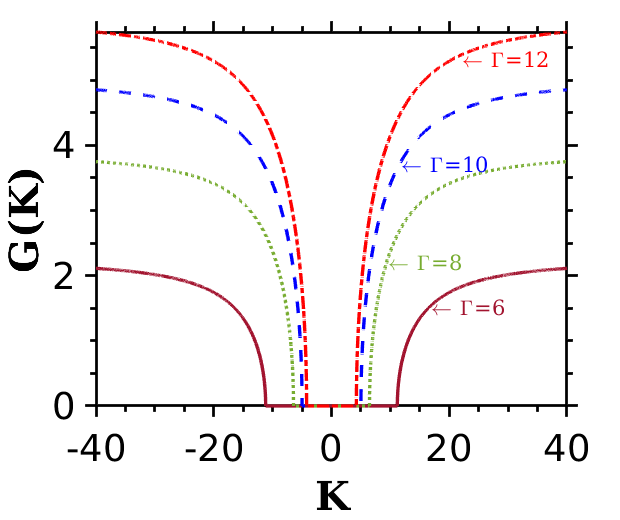}}{0.1in}{0.5in}
	\topinset{(b)}{\includegraphics[scale=0.25]{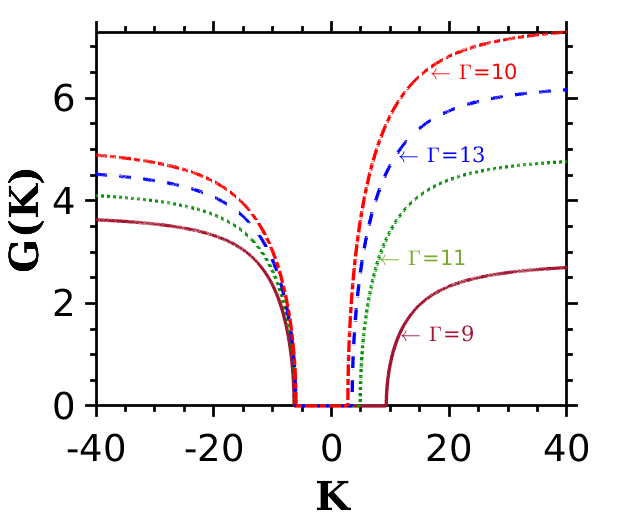}}{0.1in}{0.5in}\\
	\topinset{(c)}{\includegraphics[scale=0.24]{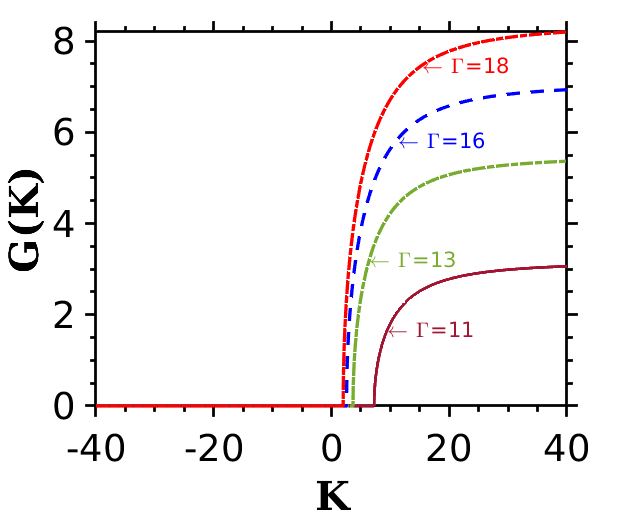}}{0.1in}{0.5in}
	\topinset{(d)}{\includegraphics[scale=0.25]{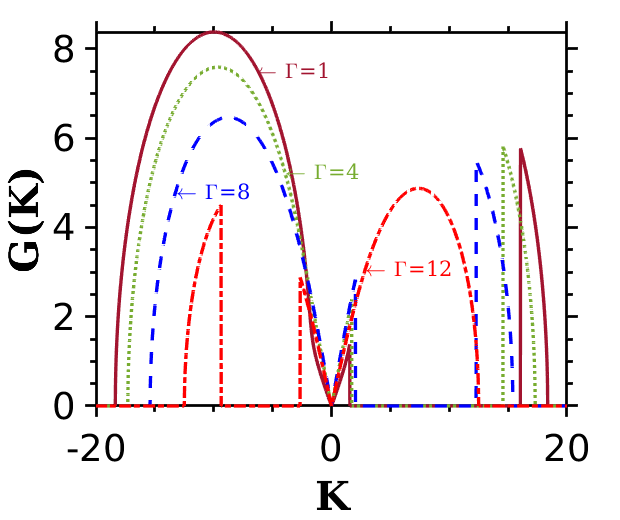}}{0.1in}{0.5in}
	\caption{(Color online) The development of instability gain spectra with the variation of the nonlinear parameter $\Gamma$ at the top of the photonic bandgap $f=1$ for four different cases. (a) conventional, (b) below,  (c) at, (d) above $\cal PT$- symmetric threshold. The parameters are assigned the values, $\kappa=5$, $P=1$, and $f=1$. }\label{Figure5}
\end{figure}

In this subsection, we analyze the impact of the nonlinear parameter $\Gamma$ on the development of MI gain spectrum at the bottom of the photonic bandgap, as presented in Fig. \ref{Figure5} with constant values of $P$ and $\kappa$ under four different cases of the $\cal PT$-symmetric system. For the conventional case, the instability spectrum as a function of $\Gamma$ can be seen in Fig.~\ref{Figure5}(a). Here, the monotonically increasing side gains are noticed on either side of the zero wavenumber $( 0)$, where the peak gain rises with an increase in the $\Gamma$ parameter. On the other hand, in the unbroken $\cal PT$- symmetric regime, as shown
in Fig. \ref{Figure5}(b), there exists the asymmetric MI  spectra, while in the $\cal PT$-symmetric threshold regime a single monotonically increasing side gain arises in the anti-Stokes wavenumber region [Fig.~\ref{Figure5}(c)]. Note that in both cases too, the growth rate tends to increase as the nonlinear saturable parameter is increased. On the other hand, the broken $\cal PT$- symmetric regime [see Fig. \ref{Figure5}(d)] reveals two distinct MI gain spectra on either side of the zero wavenumber $(K=0)$, in which the gain and bandwidth increase as $\Gamma$ is decreased, in contrast to the previous cases. Here, interestingly, when the saturation parameter is increased to $\Gamma=12$ (dot-dashed red sideband), dramatic dynamics of MI are observed where the spectra become completely shifted from Stokes to anti-Stokes wavenumber region and vice versa. 

\subsection{Instability at the top of the photonic bandgap}
\begin{figure}[t]
	\topinset{(a)}{\includegraphics[scale=0.3]{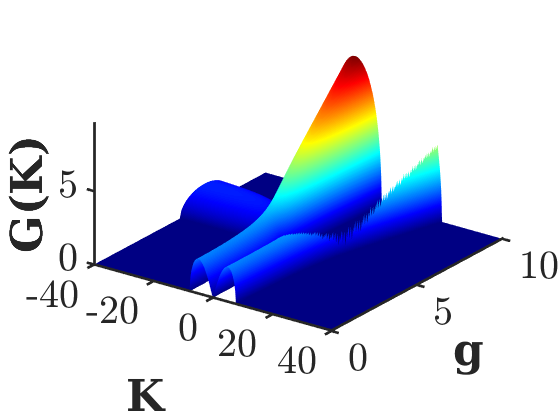}}{0.3in}{-0.45in}
	\topinset{\color{white}{(b)}}{\includegraphics[scale=0.25]{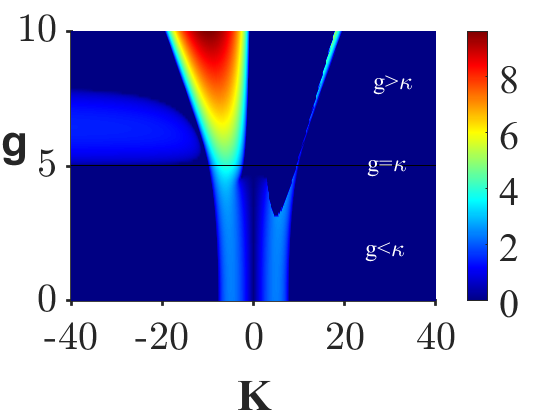}}{0.1in}{-0.45in}
\bottominset{(c)}{\includegraphics[scale=0.25]{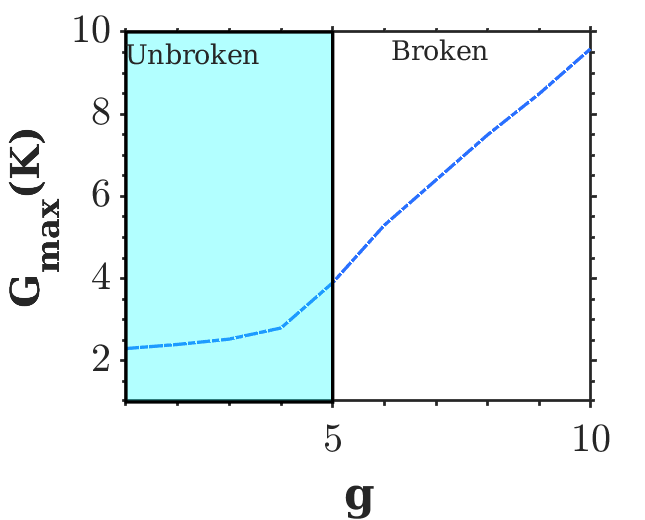}}{0.3in}{0.6in}~~
	\bottominset{(d)}{\includegraphics[scale=0.25]{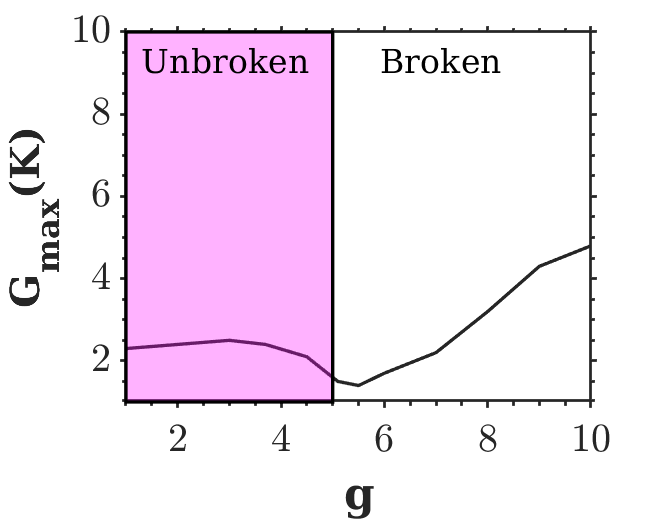}}{0.3in}{0.6in}
\topinset{(e)}{\includegraphics[scale=0.24]{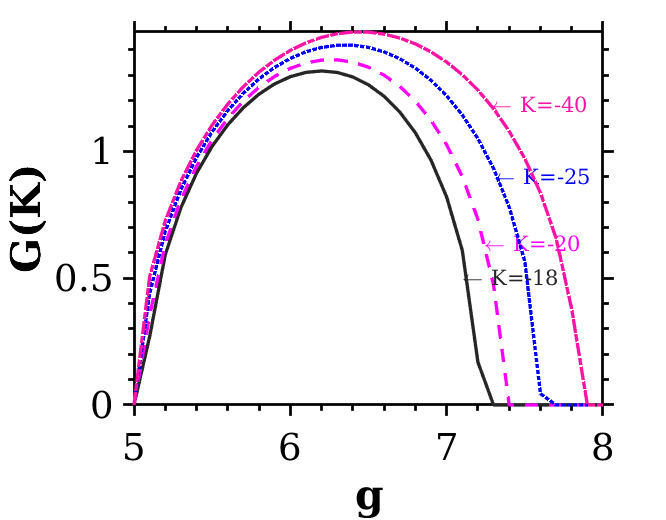}}{0.1in}{-0.4in}
	\caption{(Color online) (a) MI gain spectra as function of $g$ at the top of the photonic bandgap, $f=-1$. (b) Corresponding contour plot. The maximum gains of the MI spectra obtained in the Stokes and anti-Stokes wavenumber regions are illustrated in (c) and (d), respectively, for the variation in $g$. The peculiar MI spectrum versus the gain/loss parameter $g$ for different wavenumbers is illustrated in  (e). The parameters are assigned the values $\kappa=5$, $P=1$, and $\Gamma=3$. } \label{Figure6}
\end{figure}
In this section, we explore the dynamics of instability at the top of the photonic bandgap ($f=-1$), where the gain/loss parameter $g$ is assumed to vary continuously while the other parameters such as power, $\kappa$, and $\Gamma$ are kept constant. The instability spectrum for the top of the photonic gap is shown in  Fig.~\ref{Figure6}(a) and its corresponding contour diagram is plotted in Fig.~\ref{Figure6}(b). Here the MI gain spectrum primarily overlaps with the conventional MI gain spectrum (as obtained in the standard nonlinear Schr\"odinger equation in the anomalous dispersion regime) appearing on either side of the zero perturbation wavenumber ($K=0$) in the unbroken $\cal PT$-symmetric regime, and it continues to exhibit similar features near to the $\cal PT$-symmetric threshold region (i.e., $0<g<4.8$). Note that in this case the MI peak gain and bandwidth increase moderately with an increase in the value of $g$ on both the sides of the anti-Stokes and Stokes wavenumber until it reaches the $\cal PT$-symmetric threshold. As the value of $g$ increases further which translates the system to the broken $\cal PT$- symmetric regime, there is an emergence of a huge MI gain spectrum with a wide bandwidth in the anti-Stokes wavenumber region, while a narrow sideband emerges in the Stokes wavenumber region. Specifically, the MI spectrum found in the Stokes wavenumber region has a gain that is three times greater than the other spectrum observed in the anti-Stokes wavenumber region. Another notable ramification is the formation of a peculiar MI gain spectrum in the Stokes wavenumber region which is perpendicular to the huge primary MI spectrum. There is an increase in the value of MI gain and the bandwidth of the former as the value of $K$ increases in the negative direction. For further understanding, we separately present the growth rate of the MI gain spectrum obtained in the Stokes and anti-Stokes wave number regions in Figs.~\ref{Figure6}(c) and (d), where we observe that peak gain significantly rises when the gain/loss parameter varies from the unbroken $\cal PT$-symmetric regime to the broken $\cal PT$-symmetric regime. Following that, we present the maximum gain of the peculiar MI spectrum obtained in the Stokes wavenumber region for different values of the wavenumber $K$ in Fig.~\ref{Figure6}(e). In this case, needless to say, the maximum MI gain of the peculiar spectrum is significantly increased with increasing wavenumber $K$ besides an increase in $g$.
\subsubsection{The role of coupling coefficient ($\kappa$) }
\begin{figure}[t]
\topinset{(a)}{\includegraphics[scale=0.3]{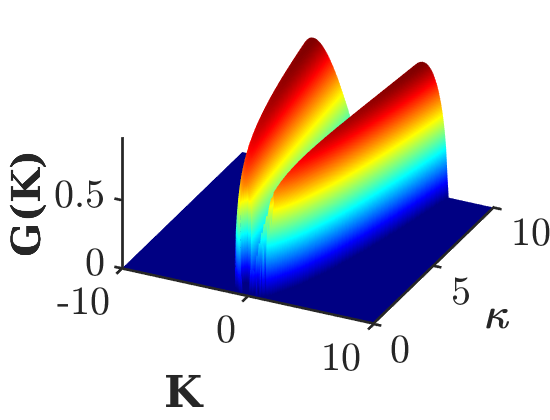}}{0.3in}{-0.4in}
	\topinset{(b)}{\includegraphics[scale=0.23]{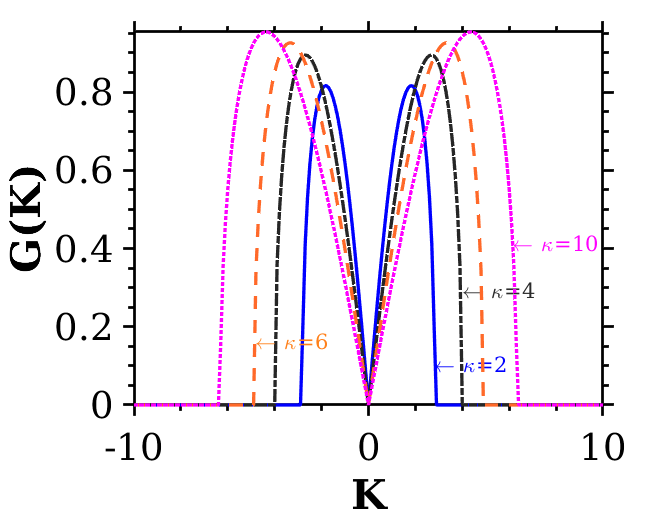}}{0.1in}{-0.3in}
 \topinset{(c)}{\includegraphics[scale=0.23]{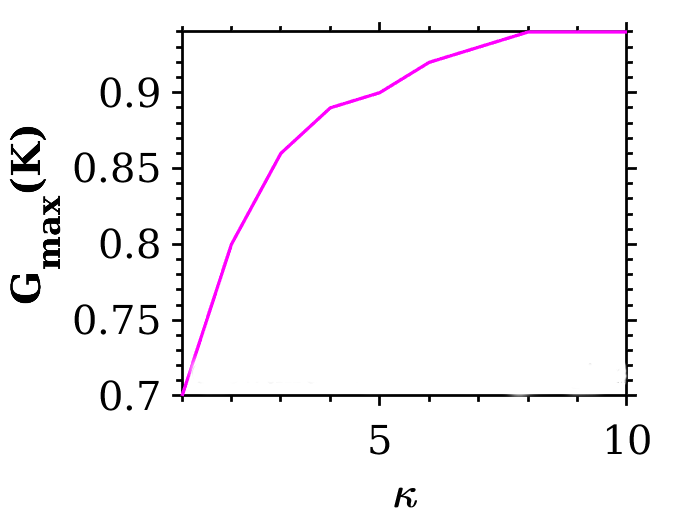}}{0.1in}{-0.3in}
	\caption{(Color online) (a) The role of coupling coefficient on the MI gain spectra at the top of the photonic bandgap $f=-1$, for the conventional case. (b) corresponding one-dimensional plot, and (c) the maximum gain of the MI spectra versus $\kappa$ in the Stokes wavenumber region.  The other parameters are fixed at $P=1$ and $\Gamma=1$. }\label{NFigure21}
\end{figure} 
We next study the MI dynamics as a function of the coupling coefficient $\kappa$ at the top of the photonic bandgap ($f=-1$), where all other parameters such as the gain/loss parameter $g$, the power $P$ and nonlinearity coefficient $\Gamma$ are kept constant with the values $P= \Gamma=1$. The conventional case is only shown here by setting $g=0$, see Fig.~\ref{NFigure21}, where one can clearly notice that the system is stable until the value of $\kappa$ attains unity. Upon increasing the value of $\kappa$, a typical MI gain spectrum appears on either side of the zero perturbation wavenumber ($K=0$), where the gain and bandwidth of the spectrum are enhanced by increasing the value of $\kappa$ further, which is then corroborated in Fig.~\ref{NFigure21}(c).
\subsubsection {Role of power }
\begin{figure}[t]
	\topinset{\color{white}{(a)}}{\includegraphics[scale=0.28]{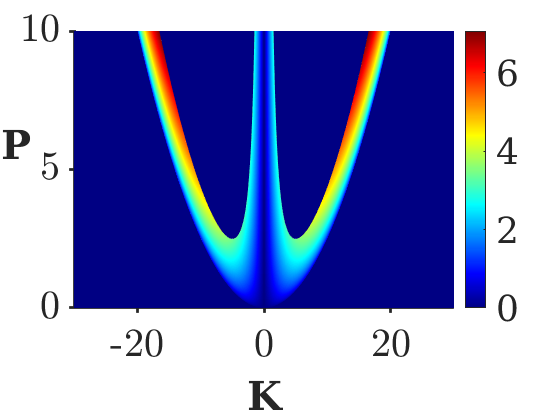}}{0.1in}{-0.4in}
	\topinset{\color{white}{(b)}}{\includegraphics[scale=0.28]{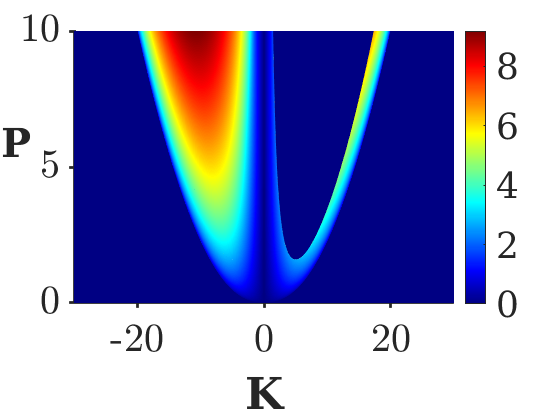}}{0.1in}{-0.4in}
	\topinset{\color{white}{(c)}}{\includegraphics[scale=0.28]{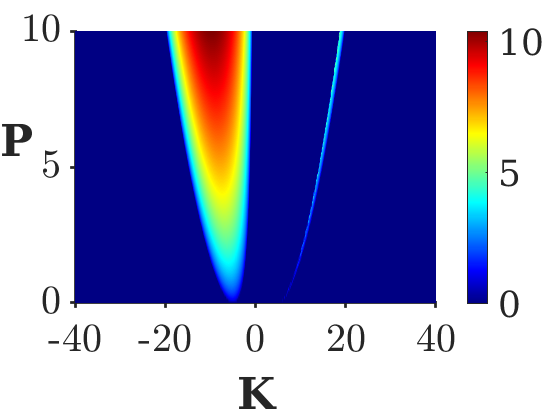}}{0.1in}{-0.4in}
	\topinset{\color{white}{(d)}}{\includegraphics[scale=0.28]{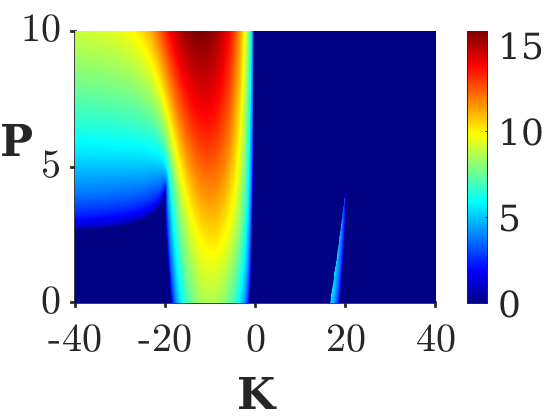}}{0.1in}{-0.4in}
	\caption{(Color online) The impact of the power on the instability spectra in the top of the photonic bandgap for (a) conventional, (b) below,  (c) at, (d) above $\cal PT$- symmetric thresholds. The parameters are assigned as $\kappa=5$, $\Gamma=2$, and $f=-1$.}\label{Figure8}
\end{figure}

To analyze the role of power at the top of the photonic bandgap, we keep the values of $\kappa$ and $\Gamma$ as $\kappa=5$ and $\Gamma=3$ with a variation in the value of $g$  that corresponds to different $\cal PT$-symmetric regimes. Figure ~\ref{Figure8}(a) shows the conventional case, where the MI gain spectrum evolves primarily as a typical MI gain spectrum on the two sides of the wavenumber. However, after a particular value of power ($P>2.7$), the MI sidebands begin to break into primary and secondary MI bands in both the Stokes and anti-Stokes wavenumber regions. In both the spectra, the gain and bandwidth of the secondary sideband are approximately twice than that of the primary sidebands and the dynamics persist even after a further increase in $P$.  It is worthwhile to mention that the structure of these MI spectra seems to have a shape of  $``\upsilon"$. In the unbroken $\cal PT$-symmetric regime, see Fig.~\ref{Figure8}(b), the previous $``\upsilon"$ shaped spectrum transforms into a complete single spectrum (without revealing any discreteness in the spectrum) in the Stokes wavenumber region, while its counterpart remains unchanged as seen in Fig.~\ref{Figure8}(a). The impact of the power in the $\cal PT$-symmetric threshold region is depicted in Fig. \ref{Figure8}(c). In this case, the MI spectrum in the Stokes wavenumber region gets enhanced moderately in both its gain and width. On the other hand, the spectrum observed in the anti-Stokes wavenumber region is drastically suppressed revealing a very thin bandwidth. In the broken $\cal PT$-symmetric regime (see Fig.~\ref{Figure8}(d)), when the value of $P$ is low, there exists an emergence of the typical spectrum after which one can observe the emergence of an additional peculiar MI gain spectrum in the Stokes wavenumber region with an increase in the value of $P$ ($P>3.5$). On the other hand, the sideband on the anti-Stokes wavenumber region almost disappears compared to the previous case. Note that in all of the above four cases, it is clear to notice a common rise in the gain and bandwidth of the MI spectra as the value of $P$ increases.
\subsubsection {Impact of nonlinear parameter  $\Gamma$ }
\begin{figure}[t]
	\topinset{(a)}{\includegraphics[scale=0.25]{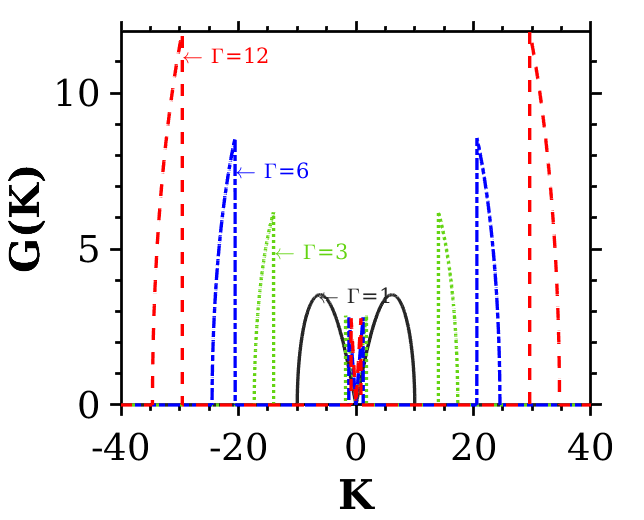}}{0.1in}{-0.45in}
	\topinset{(b)}{\includegraphics[scale=0.25]{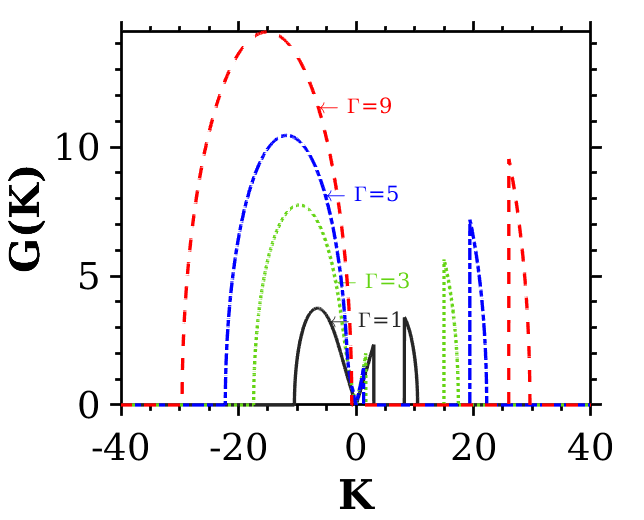}}{0.1in}{-0.45in}
	\topinset{(c)}{\includegraphics[scale=0.25]{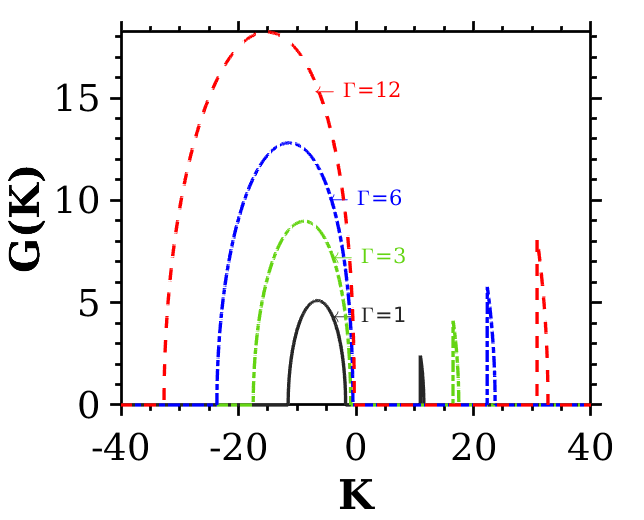}}{0.1in}{-0.45in}
	\topinset{(d)}{\includegraphics[scale=0.25]{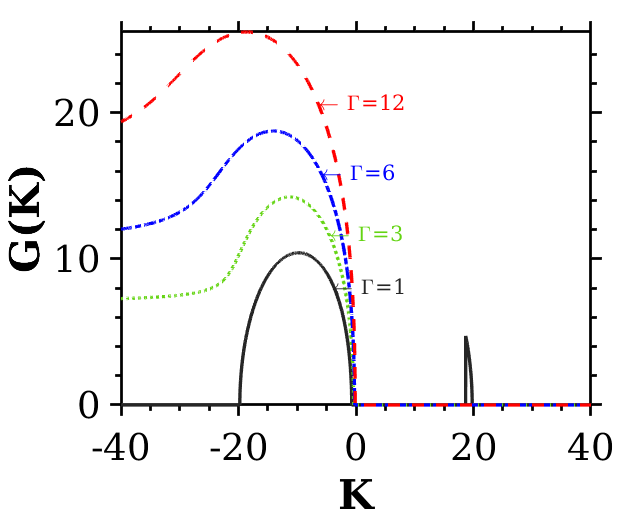}}{0.1in}{-0.45in}
	\caption{(Color online) The role of nonlinear $\Gamma$ parameter on the MI gain spectra in the top photonic bandgap for (a) conventional, (b) below,  (c) at, (d) above the $\cal PT$- symmetric threshold. The parameters are assigned the values $\kappa=5$, $P=5$, and $f=-1$. }\label{Figure9}
\end{figure}

We now examine the impact of the nonlinear parameter $\Gamma$ at the top of the photonic bandgap in different regimes of $\cal PT$ symmetry. These results are shown in Fig.~\ref{Figure9}, with fixed values of system parameters, $P$ and $\kappa$. In the conventional case, the spectrum primarily exhibits a typical MI gain spectrum for the low value of the nonlinear parameter ($\Gamma=1$), as shown in Figs.~\ref{Figure9}(a). However, as $\Gamma$ increases, the secondary MI spectrum starts to emerge followed by the former primary MI gain on either side of the zero wavenumber. In addition, as the value of $\Gamma$ increases, the spectrum gets enhanced and the separation distance between the primary and secondary MI bands also increases by shifting towards the higher wavenumber. We then analyze the MI dynamics in the unbroken $\cal PT$-symmetric regime, which is illustrated in Fig.~\ref{Figure9}(b). Here it is obvious to note that the system exhibits asymmetric MI gain spectra on both sides of the wavenumber including a huge primary MI spectrum in the Stokes wavenumber region while its counterparts remain almost unchanged except for $\Gamma=1$ as in Fig.~\ref{Figure9}(a). Further, it can be seen in Fig.~\ref{Figure9}(c) that the MI gain spectrum retains the same dynamics even when moving to the $\cal PT$-symmetric threshold regime. Nevertheless, for each nonlinear saturation parameter value, the peak gain of the sidebands in the Stokes wavenumber region rises moderately compared to the previous cases shown in Figs.~\ref{Figure9}(a) and (b). But the gain in the sidebands on the other side falls as the value of $\Gamma$ increases. In the broken $\cal PT$-symmetric regime, the sidebands become a monotonically increasing gain in the Stokes wavenumber region, while the other side displays the stable dynamics of CW state except for $\Gamma=1$, as shown Fig.~\ref{Figure9}(d).
\subsection {Modulational instability in the anomalous dispersion regime}
\begin{figure}[t]
	\topinset{(a)}{\includegraphics[scale=0.28]{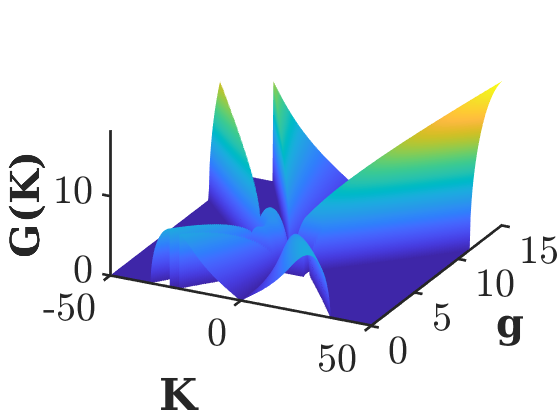}}{0.4in}{-0.4in}
	\topinset{(b)}{\includegraphics[scale=0.28]{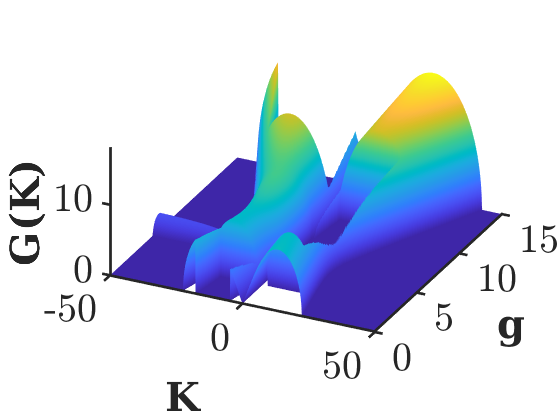}}{0.4in}{-0.4in}
	\topinset{(c)}{\includegraphics[scale=0.28]{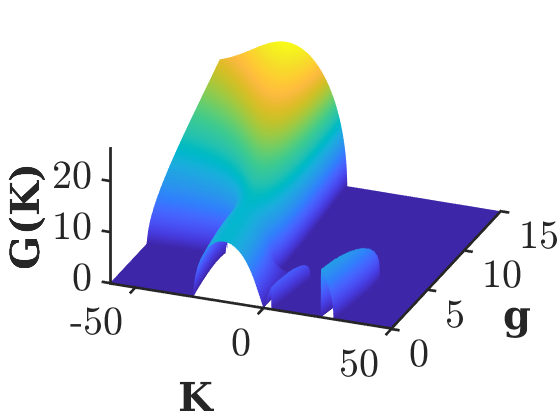}}{0.4in}{-0.4in}
	\topinset{(d)}{\includegraphics[scale=0.28]{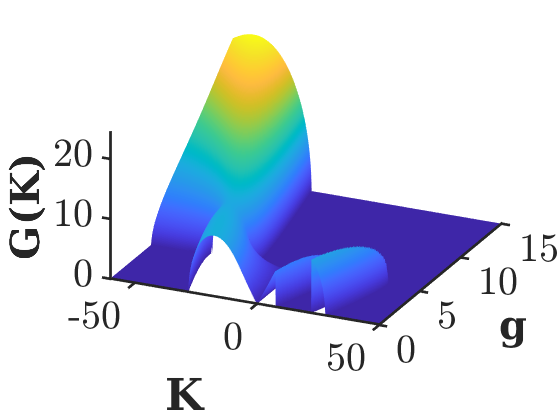}}{0.4in}{-0.4in}
	\caption{(Color online) Instability gain spectra in the anomalous dispersion regime as a function of $g$ for different dispersion values, including (a) $f=-0.1$, (b) $f=-0.3$, (c) $f=-3$, and (d) $f=-5$. The other parameters are fixed as $P=10$, $\kappa=5$, and $\Gamma=3$.}\label{Figure10}
\end{figure}

Another significant part of this study is to investigate the dynamics of MI in the anomalous dispersion regime of the proposed system \eqref{PTCoupler}. As studied in the previous sections, we here also investigate the emergence of MI gain spectrum under three $\cal PT$-symmetric regimes, besides the conventional case.  Also, we investigate the MI gain spectrum for four different dispersion parameter values in the anomalous dispersion regime. Compared to the earlier results presented in the paper, all of these cases give rise to various unique MI gain spectra. For instance, in the unbroken $\cal PT$-symmetric regime, two different MI sidebands can be observed when the value of $f$ is fixed at  $f=-0.1$, see Fig. \ref{Figure10}(a). It is to be noted that the MI sideband located in the anti-Stokes wavenumber region side is slightly larger in gain and wider in bandwidth, as compared to the spectrum that appeared on the other side. It is also observed that an additional secondary MI spectrum emerges closer to the MI sideband in the Stokes wavenumber region. When we increase the value of the gain/loss parameter further, in particular, towards $\cal PT$-symmetric threshold, all these MI spectra tend to merge in the Stokes wavenumber region and disappear in the anti-Stokes wavenumber region due to the presence of singularity. 
In the broken $\cal PT$-symmetric threshold regime,  one can notice the magnified mirror image of the MI gain spectrum found in the unbroken $\cal PT$-symmetric regime. However, the gain and bandwidth of the MI spectrum are significantly enhanced in this case compared to the unbroken regime.

Figure \ref{Figure10}(b) shows the MI gain spectrum as a function of $g$ 
by fixing the dispersion parameter as $f=-0.5$. In this case, the patterns of these MI sidebands seem to be more complicated than the previous one with some irregularities in all the $\cal PT$-symmetric regimes. To elucidate the dynamics further, it is apparent that most of the MI dynamics resemble the previous spectra except for the emergence of an additional peculiar MI spectrum parallel to the primary MI spectrum formed in the anti-Stokes wavenumber region. This spectrum continues to extend as the wavenumber increases. By further decreasing the value of the dispersion parameter to $f=-3$, it becomes evident that the dynamics of the MI gain spectrum, portrayed in Fig.~\ref{Figure10}(c), is significantly simpler than in the previously studied cases (cf.  Figs.~\ref{Figure10}(a) and \ref{Figure10}(b)). There are two distinct MI sidebands, where the gain of the MI sideband in the Stokes wavenumber region is comparatively higher than the spectrum on the anti-Stokes wavenumber region. Additionally, a secondary MI band can be observed in the anti-Stokes wavenumber region with a higher MI gain than the primary MI spectrum. Upon increasing the value of $g$ further, the MI sideband in the Stokes wavenumber region transforms into a pronounced monotonically increasing gain. The structure of the MI spectrum remains the same when the dispersion parameter is further decreased to $f=-5$, as illustrated in Fig.~\ref{Figure10}(d). However, the gain of the primary and secondary MI bands is significantly enhanced from the spectra obtained when $f=-3$.

\subsubsection{Influence of $\kappa$ on the MI gain spectrum }
\begin{figure}[t]
	\topinset{(a)}{\includegraphics[scale=0.3]{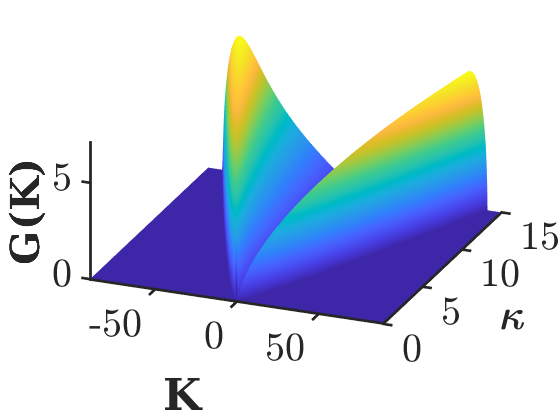}}{0.4in}{-0.4in}
	\topinset{(b)}{\includegraphics[scale=0.23]{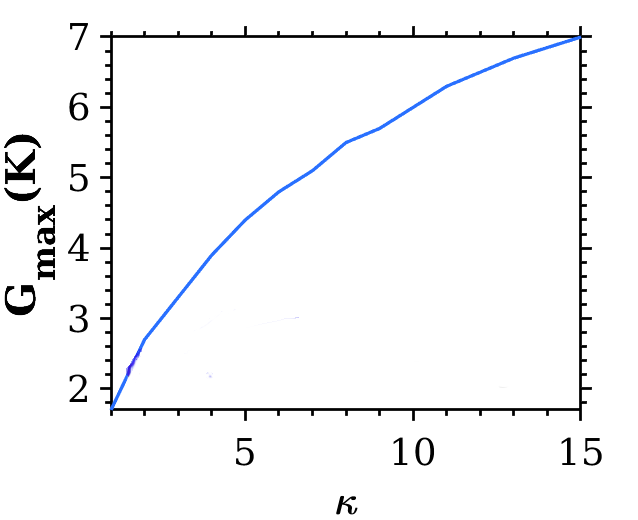}}{0.1in}{-0.4in}
	\caption{(Color online) The MI gain spectra as a function of $\kappa$ for the anomalous dispersion regime. (b) The maximum gain of the peculiar spectra versus $\kappa$ in the Stokes wavenumber region. The parameters are assigned as $g=0$, $\Gamma=1$, $P=10$ and $f=-0.1$. }\label{NFigure3}
\end{figure}

Figure \ref{NFigure3} shows the impact of the coupling coefficient ($\kappa$) on the formation of MI gain spectrum in the anomalous dispersion regime for $f=-0.1$ in the conventional case alone. For this purpose, we continuously change the coupling coefficient $\kappa$ while maintaining constant values for the gain/loss parameter, power, and saturable nonlinear parameter. As in the previous case, shown in Fig.~\ref{NFigure21}(a), here too, a typical MI gain spectrum is observed which further extends on either side of the zero perturbation wavenumber as the coupling coefficient varies.  When $\kappa$ is increased, the MI gain and bandwidth of the spectrum are significantly increased in the anti-Stokes and Stokes wavenumber regions, which is further corroborated in Fig. \ref{NFigure21}(b).
\subsubsection {Influence of input power ($P$) }
\begin{figure}[t]
	\topinset{\color{white}{(a)}}{\includegraphics[scale=0.28]{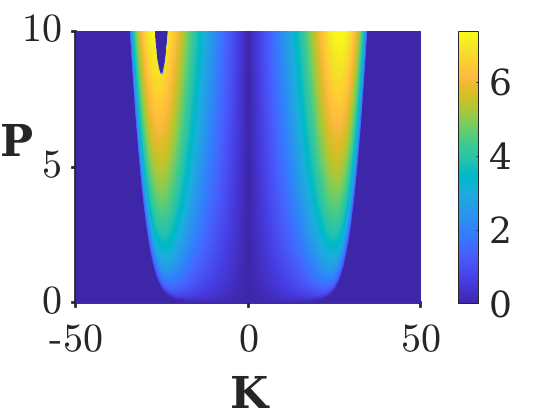}}{0.1in}{-0.4in}
	\topinset{\color{white}{(b)}}{\includegraphics[scale=0.28]{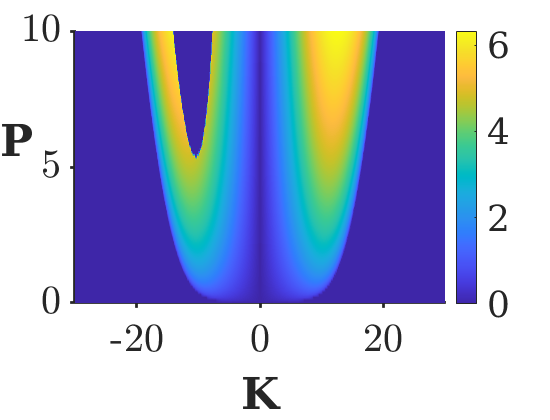}}{0.1in}{-0.4in}
	\topinset{\color{white}{(c)}}{\includegraphics[scale=0.28]{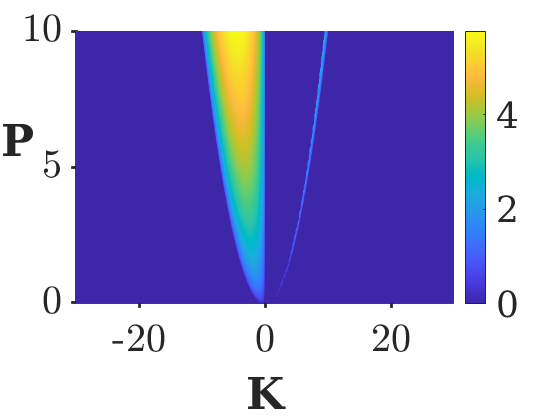}}{0.1in}{-0.45in}
	\topinset{\color{white}{(d)}}{\includegraphics[scale=0.28]{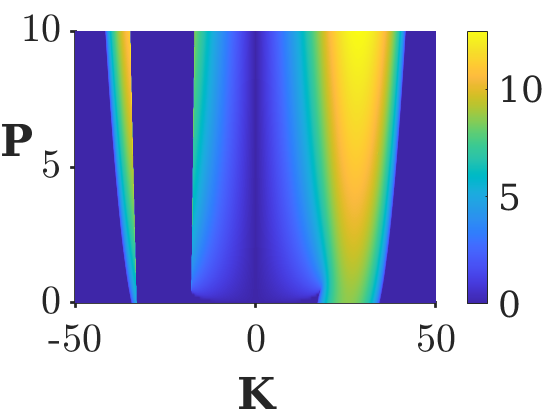}}{0.1in}{-0.45in}
	\caption{(Color online) The MI gain spectra in the anomalous dispersion regime, as a function of $P$ for (a) conventional, (b) below, (c) at, and  (d) above $\cal PT$- symmetric thresholds. The parameters are assigned as $\kappa=5$, and $\Gamma=3$, and $f=-0.1$. }\label{Figure11}
\end{figure}

The role of input power in the anomalous dispersion regime has also been analyzed in greater detail for all the  $\cal PT$-symmetric threshold regimes as well as the conventional case. Figure~\ref{Figure11}(a) delineates the MI gain spectrum for the conventional case as a function of power. Here, the typical MI gain spectrum is observed around the zero perturbation wavenumber and the value of gain and bandwidth become more pronounced by further increasing the value of $P$. In particular, one can witness a small range of irregularities due to the singularity arising in the Stokes wavenumber region. The same MI structure remains almost unchanged qualitatively when the system is operated in the unbroken $\cal PT$-symmetric regime, see Fig.~\ref{Figure11}(b). However, the gain and bandwidth of the MI spectrum are somewhat reduced and the range of irregularities that appear in the Stokes wavenumber region is extended.
The scenario has been changed when switching to the $\cal PT$-symmetric threshold regime, where one can observe the manifestation of asymmetric MI gain spectra that include the typical MI spectrum in the Stokes wave number region and a very narrow side band in the anti-Stokes wavenumber region, as shown in Fig.~\ref{Figure11}(c). In this case, too the MI gain increases with the increase in the value of $P$. In the broken $\cal PT$-symmetric regime, see Fig.~\ref{Figure11}(d), there exist two asymmetric MI spectra, which include a wider MI spectrum in the anti-Stokes wavenumber region and two different MI sidebands in the Stokes wavenumber side.

\subsubsection {Impact of nonlinear saturation parameter $\Gamma$ }
\begin{figure}[t]
	\topinset{(a)}{\includegraphics[scale=0.25]{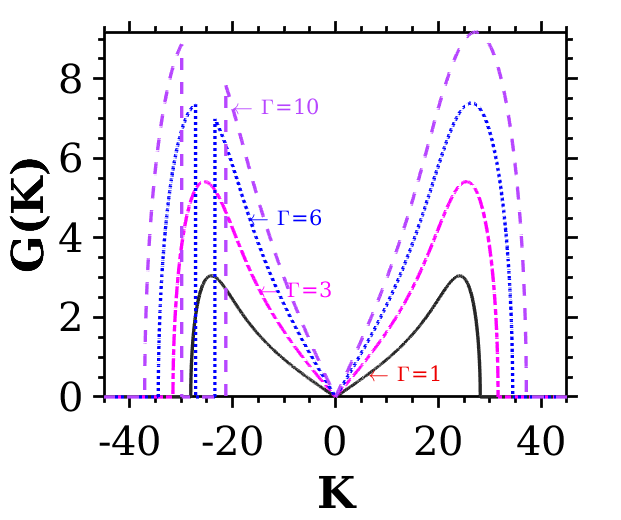}}{0.1in}{-0.45in}
	\topinset{(b)}{\includegraphics[scale=0.25]{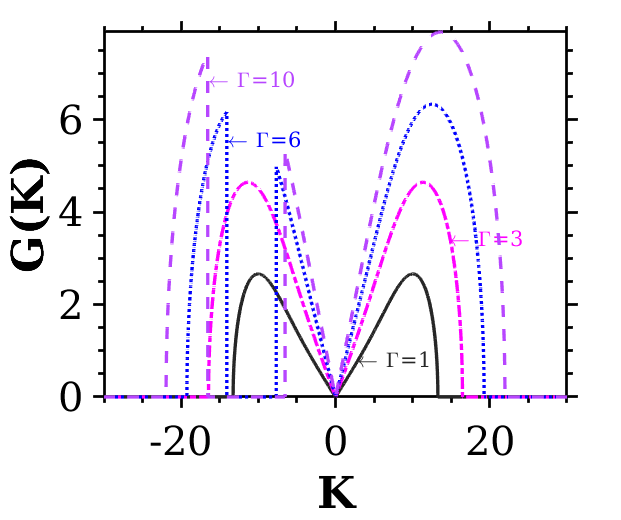}}{0.1in}{-0.45in}
	\topinset{(c)}{\includegraphics[scale=0.25]{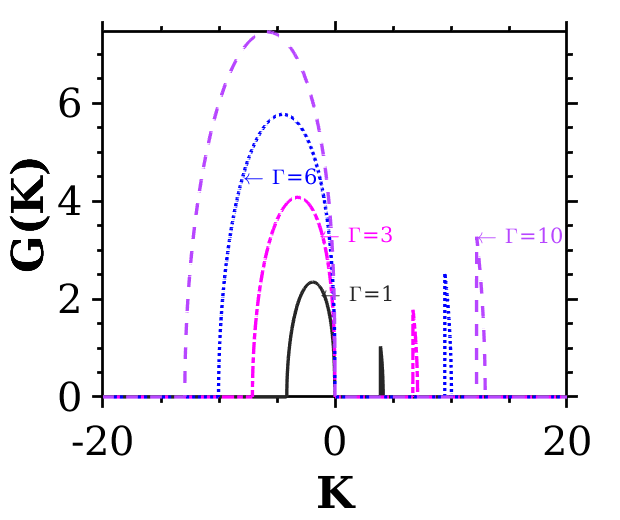}}{0.1in}{-0.45in}
	\topinset{(d)}{\includegraphics[scale=0.26]{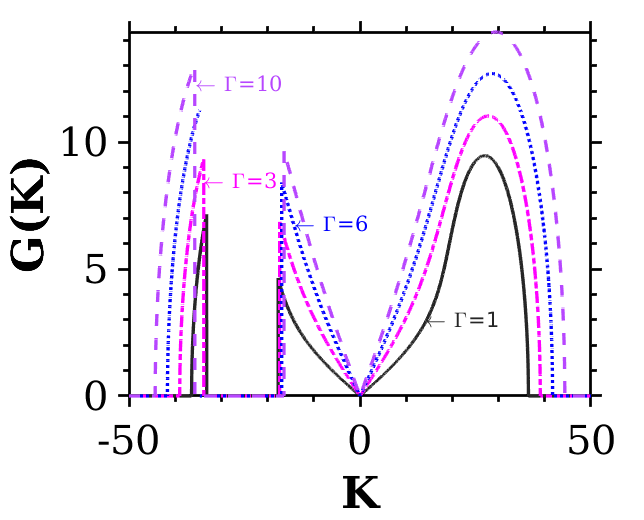}}{0.1in}{-0.45in}
	\caption{(Color online) The MI gain spectra as a function of $\Gamma$ in the anomalous dispersion regime, for (a) conventional, (b) below, (c) at, and (d) above $\cal PT$- symmetric thresholds with, $f=-0.1$ and $\kappa=P=5$. }\label{Figure12}
\end{figure}

We now investigate the influence of the nonlinear saturation parameter on the instability spectrum in the anomalous dispersion regime under three different $\cal PT$-symmetric conditions with the conventional case. Figure \ref{Figure12}(a) shows the MI dynamics in the conventional case where it is observed a symmetric MI spectrum on both sides of the wavenumber. However, when the value of the saturation parameter is increased to $\Gamma=6$, the spectrum becomes asymmetric, wherein two different MI bands start to appear. Also, the MI gain and bandwidth of both the primary and secondary MI gain spectra increase as the value of $\Gamma$ increases. When the system is operated in the unbroken regime (see Fig. \ref{Figure12}(b)), though the MI spectrum seems to overlap with the conventional case for lower values of  $\Gamma$, a further increase in the value of $\Gamma$, for instance, $\Gamma=6$, the symmetric spectrum transforms into an asymmetric one. What differentiates this from the former is that the spectrum results in a slightly narrower bandwidth and lower gain compared to the conventional case. The exceptional point, shown in Fig. \ref{Figure12}(c), reveals a different MI spectrum where very thin and lower gain MI peaks emerge in the anti-Stokes wavenumber region while on the other side, the spectrum features a wider bandwidth and higher gain.  Finally, in the broken $\cal PT$-symmetric case as illustrated in Fig.~\ref{Figure12}(d), it retains the same MI patterns obtained in both the conventional and unbroken $\cal PT$-symmetric cases. Nevertheless, the separation distance between the primary and secondary MI bands is much higher in the Stokes wavenumber region while the spectrum in the anti-Stokes wavenumber region exhibits a much wider bandwidth.

\subsection {Modulational instability in the normal dispersion regime}
\begin{figure}[t]
	\topinset{(a)}{\includegraphics[scale=0.28]{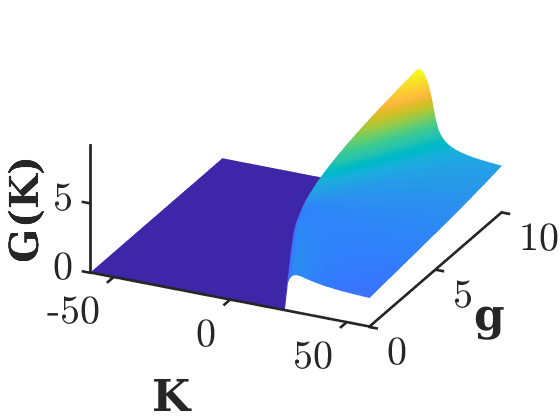}}{0.4in}{-0.4in}
	\topinset{(b)}{\includegraphics[scale=0.28]{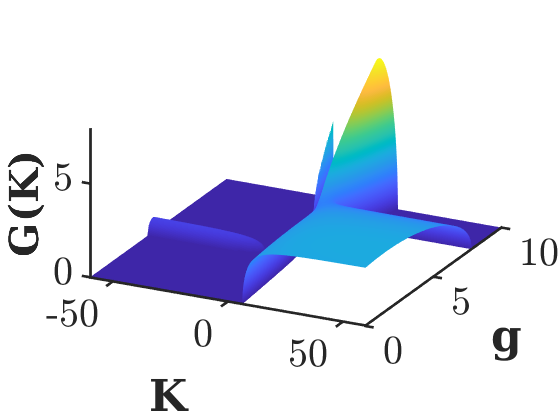}}{0.4in}{-0.4in}
	\topinset{(c)}{\includegraphics[scale=0.28]{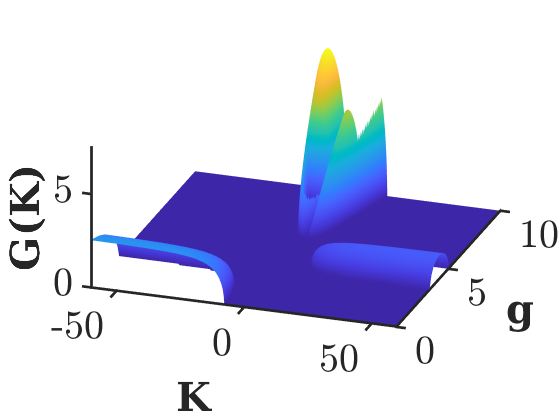}}{0.4in}{-0.4in}
	\topinset{(d)}{\includegraphics[scale=0.28]{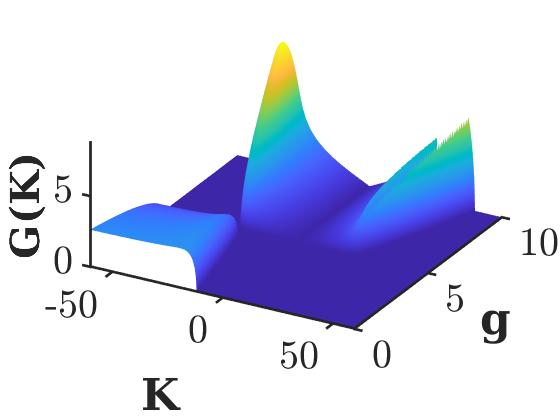}}{0.4in}{-0.4in}
	\caption{(Color online) The MI gain spectra as a function of gain/loss parameter in the normal dispersion regime for different $f$ values, including (a) $f=0.1$, (b) $f=0.5$, (c) $f=0.7$, and (d) $f=5$. Here, the other parameters are fixed as $\kappa=5$ and $P=\Gamma=2$.}\label{Figure13}
\end{figure}

This section brings another important study of the MI spectrum obtained in the normal dispersion regime $(f>0)$. For this reason, all the parameters except $g$ are fixed, and like in the previous section,  we present the investigations by assigning different values of the dispersion parameter $f$ by continuously varying the gain/loss parameter.  The MI gain spectrum is shown in Fig.~\ref{Figure13}(a) as a function of $g$ when the value of $f$ is set to a low value, such as $f=0.1$. Unlike in the anomalous dispersion regime, here, the MI spectrum takes place only in the anti-Stokes wavenumber region and on further increasing the value of gain/loss coefficient, the spectrum drifts towards the zero perturbation wavenumber with a magnitude drop in its gain. Near the broken $\cal PT$-symmetric region, the MI spectrum in the anti-Stokes wavenumber region re-drifts in the opposite direction. 

Upon further increasing the value of $f$ to 0.5 ($f=0.5$) (Fig. \ref{Figure13}(b)), a unique MI gain spectrum has been observed. The spectrum primarily appears with a monotonically increasing sideband gain in the anti-Stokes wavenumber region. In parallel, a peculiar MI gain spectrum can be seen in the Stokes wavenumber region, which is more enhanced as $K$ increases. Once the system reaches $\cal PT$-symmetry broken regime, the monotonously increasing  MI side gain transforms into two distinct primary MI spectra around $K=0$, with the MI sideband in the anti-Stokes wavenumber regime being more prominent than the MI sideband on the other side, whose gain and bandwidth increase with increasing $g$ values. We notice that the scenario has been changed a bit when we assign the value of $f$ as $f=0.7$. This increase in the value of the dispersion parameter separates the monotonically increasing spectrum from the primary MI spectrum in the anti-Stokes wavenumber regime, as shown in Fig.~\ref{Figure13}(c). Also, the peculiar MI spectrum in the negative wavenumber region is significantly enhanced as compared to the previous case (see Fig.~\ref{Figure13}(b)). 
In contrast to the previous spectrum shown in Fig. \ref{Figure13}(b), the broken $\cal PT$-symmetric regime produces a comparatively wide and large gain spectrum.  For $f=5$, the system reveals quite a different MI structure as compared to the previous case ( $f=0.7$), in which the peculiar MI gain spectrum disappears in the anti-Stokes wavenumber regime and the monotonically increasing side gain shifts from anti-Stokes to Stokes wave number region. Further, in the broken $\cal PT$-symmetric regime, the primary MI gain spectra emerge on either side of the wavenumber with pronounced gain and bandwidth, as shown Fig.~\ref{Figure13}(d). 
\subsubsection{Role of $\kappa$ on the MI gain spectrum}
\begin{figure}[t]
	\topinset{(a)}{\includegraphics[scale=0.3]{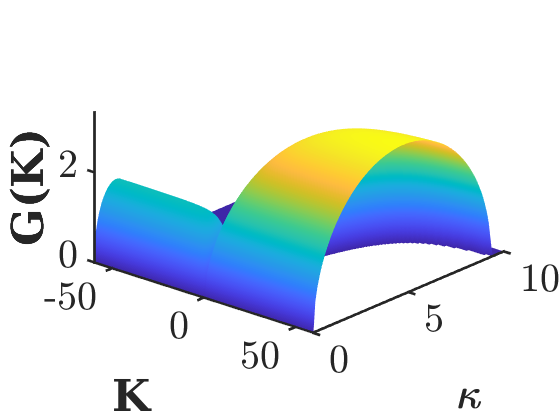}}{0.4in}{-0.4in}
	\topinset{(b)}{\includegraphics[scale=0.23]{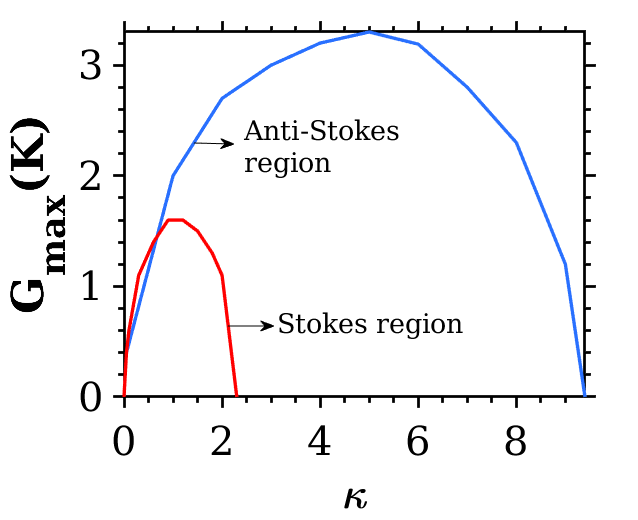}}{0.1in}{-0.4in}
	\caption{(Color online) The MI gain spectra as a function of the coupling coefficient $\kappa$ in the normal dispersion regime. (b) The maximum gain of the peculiar MI gain as a function of coupling coefficient $\kappa$. Here, the parameters are $g=0$, $P=5$, and $\Gamma=1$.}\label{Nfigure4}
\end{figure}

In this subsection, we analyze the impact of the coupling coefficient $\kappa$ on the MI gain spectrum in the normal dispersion regime for $f=0.7$, in the conventional case alone by setting the gain/loss parameter to zero as shown in Fig.~\ref{Nfigure4}(a). Interestingly, it shows two different peculiar MI gain spectra that appear around the zero wavenumber ($K=0$), where the peculiar MI gain spectrum observed in the anti-Stokes wavenumber regime feature a wider bandwidth compared to the other one found in the Stokes wavenumber region. In particular, the gain of this peculiar MI  spectrum is twice as high as in the MI spectrum on the other side, while the bandwidth is about four times wider. It should be stressed that such an unusual MI gain spectrum as a result of the coupling coefficient is a new finding in the context of coupled nonlinear systems. The peak gain traced as a function of $\kappa$, displayed in  Fig.~\ref{Nfigure4}(b), also confirms the same.
\subsubsection {Influence of input power $P$ }
\begin{figure}[t]
	\topinset{(a)}{\includegraphics[scale=0.25]{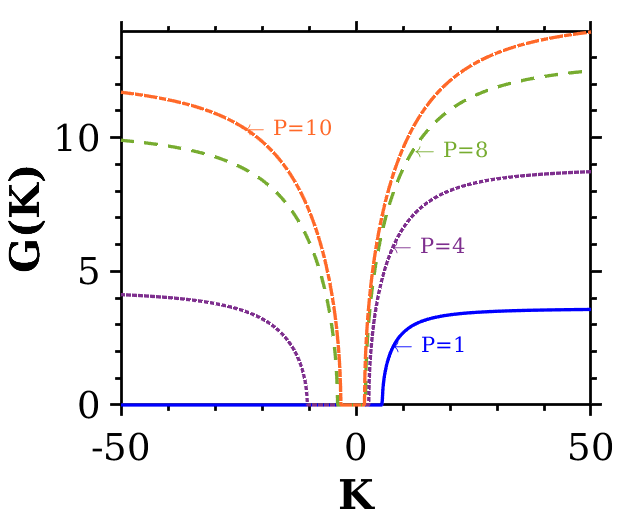}}{0.1in}{-0.45in}
	\topinset{(b)}{\includegraphics[scale=0.25]{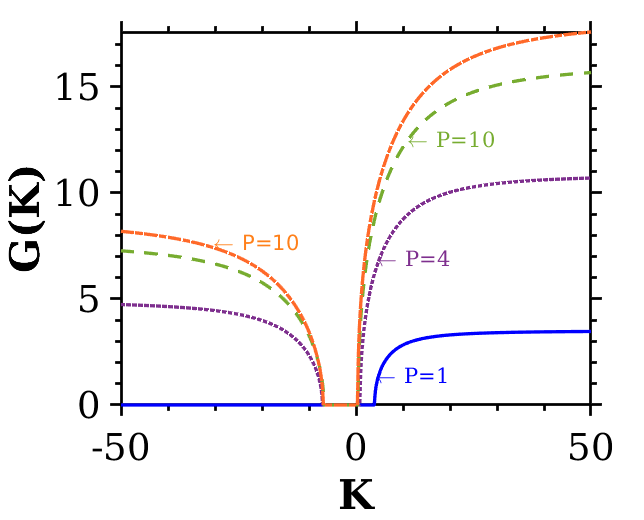}}{0.1in}{-0.45in}
\topinset{(c)}{\includegraphics[scale=0.25]{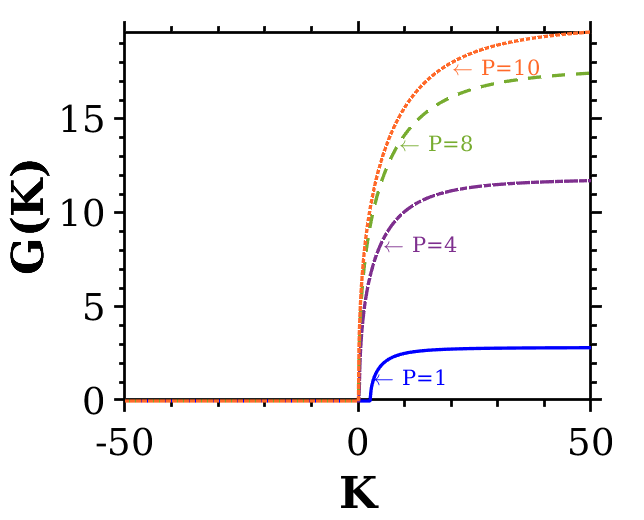}}{0.1in}{-0.45in}
	\topinset{(d)}{\includegraphics[scale=0.25]{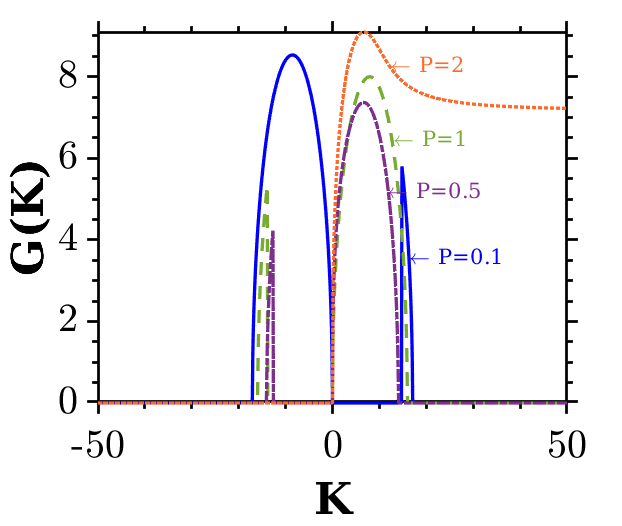}}{0.1in}{-0.45in}
	\caption{(Color online) The MI gain spectra as a function of input power in the normal dispersion regime, for (a) conventional, (b) below,  (c) at, (d) above $\cal PT$-thresholds. The parameters are assigned with the values $\kappa=5$, $\Gamma=5$, and $f=0.6$.}\label{Figure14}
\end{figure}

We next analyze the role of power on the characteristics of the instability spectrum in the normal dispersion regime under four different $\cal PT$-symmetric regimes in Fig.~\ref{Figure14}. First in the conventional case (Fig. \ref{Figure14}(a)), one can observe a monotonically increasing MI gain only in the anti-Stokes wavenumber region for $P=1$. However, when we increase the value of input power further, the monotonically increasing gain appears on both sides of the wavenumber. Note that the peak gain in the anti-Stokes wavenumber is higher than the one in the Stokes wavenumber region. The MI characteristics retain the same dynamics and pattern when moving on to the unbroken $\cal PT$- symmetric regime, as shown in Fig. \ref{Figure14}(b).  However, the gain value of the monotonically increasing gain located in the Stokes wavenumber region has been considerably suppressed. When it comes to the case of $\cal PT$-symmetric threshold, the monotonically increasing gain in the Stokes wavenumber region is completely suppressed while the spectrum found on the other side remains unchanged, which is shown in Fig. \ref{Figure14}(c). For the case of broken $\cal PT$-symmetric regime, see Fig. \ref{Figure14}(d), the MI spectrum completely transforms into a typical one for the lower value of the power $P=0.1$. With a further increase in the value of $P$, for example, $P=0.5$, and $P=1$, the spectra get swapped on the different wavenumber regions.  When we increase the value of $P$ to a higher value such as $P=2$, the MI sideband located in the anti-Stokes wavenumber regime transforms into a monotonically increasing gain, while the MI spectrum on the other side vanishes.
\subsubsection {Impact of nonlinear saturation parameter ($\Gamma$)}
\begin{figure}[t]
	\topinset{\color{white}{(a)}}{\includegraphics[scale=0.25]{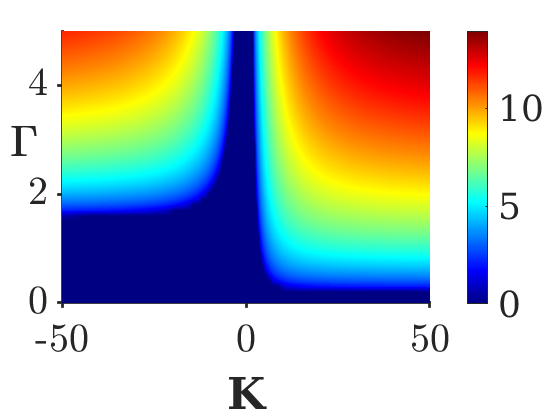}}{0.1in}{-0.4in}
	\topinset{\color{white}{(b)}}{\includegraphics[scale=0.25]{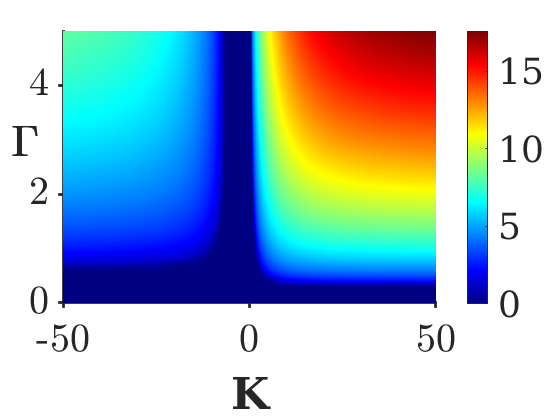}}{0.1in}{-0.4in}
	\topinset{\color{white}{(c)}}{\includegraphics[scale=0.25]{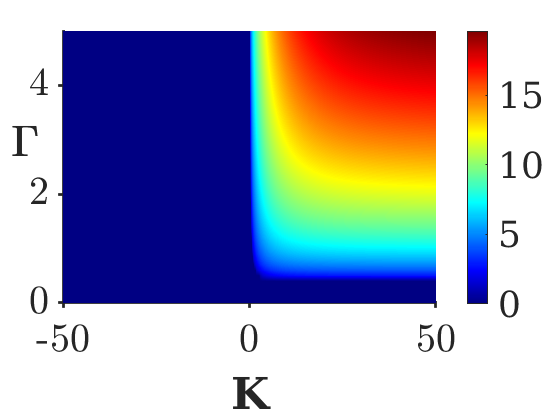}}{0.1in}{-0.4in}
	\topinset{\color{white}{(d)}}{\includegraphics[scale=0.25]{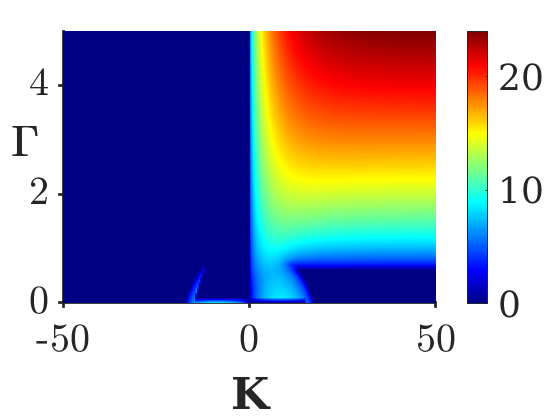}}{0.1in}{-0.4in}
	\caption{(Color online) The MI gain spectra with the variations of $\Gamma$ in the normal dispersion regime, for (a) conventional, (b) below,  (c) at, (d) above $\cal PT$-thresholds. The parameters are assigned as $\kappa=5$, and $P=10$, and $f=0.6$. }\label{Figure15}
\end{figure}

We finally investigate the effect of the nonlinear saturation parameter on the MI gain spectrum in the normal dispersion under four different conditions, including the conventional case as shown in Fig. \ref{Figure15}. When $g=0$ (conventional case), see Fig.~\ref{Figure15}(a), it produces an asymmetric MI structure consisting of two different monotonically increasing gains, where the peak gain in the anti-Stokes wavenumber region is more pronounced than the other side. Figure \ref{Figure15}(b) depicts the MI dynamics in the unbroken $\cal PT$-symmetric regime, where one can observe that the system almost retains the same MI patterns though the range of monotonically increasing gain gets increased in the Stokes wavenumber region. 
Further,  when the system is operated at the $\cal PT$-symmetric threshold, the monotonically increasing gain disappears in the Stokes wavenumber region as shown in Fig. \ref{Figure15}(c). In contrast to the above, the broken $\cal PT$-symmetric regime reveals a narrow side-band in both the Stokes wavenumber regions for small values of $\Gamma$, as can be seen in Fig.~\ref{Figure15}(d). Comparing the gain of all these spectra, it is to be noted that the broken $\cal PT$ - symmetric produces a relatively higher gain. As the system exhibits rich and complex MI patterns,  for an easier and better understanding of the ramifications obtained under various cases, we have provided the summary of the main results in a tabular form in Tab. 1 when the system changes from the conventional to different $\cal PT$- symmetric regimes as a function of $g$.

\begin{table*}
\renewcommand{\arraystretch}{1.2}
\caption{A Summary of the MI gain spectrum obtained in the PT-FBG with saturable nonlinearity}
		\begin{tabular}{l llllllll}
			\hline\\
			& \multicolumn{4}{c}{Does the MI gain spectrum exist?} & \multicolumn{3}{c}{Types of MI gain spectrum} \\\\
			\cline{2-5} \cline{6-9}\\
			Types of   &Convl   & Unbroken & Exceptional& Broken &Convl& Unbroken   & Exceptional  & Broken  \\
		regimes&case	& regime   & point  &  regime &case& regime   & point  & regime  \\\\\hline\\
		Top   & No   & Yes & No & Yes   &Nil& (i) Type-IV MI & Nil&(i) Type-II MI \\
		 photonic &	   & 	& 	& 	 &  &(ii) Type-V MI  & &(ii) Type-III MI \\
		 bandgap &	   & 	& 	& 	 &  & & & \\\hline\\
			Bottom &Yes & Yes & Yes & Yes   & Type-I MI &Type-I MI &Type-II MI & (i) Type-II MI \\
			photonic& &  &  &   &  & & &(ii)  Type-III MI \\
			bandgap& &  &  &   &  & & &(iii) Type-IV MI \\\\\hline\\
			Anamalous  &  Yes  & Yes & Yes & Yes  &  (i) Type-II MI& (i)Type-II MI & (i) Type-V MI & (i) Type-II MI   \\
			dispersion&&  &  &   & (ii) Type-III MI  &(ii) Type-III MI  &  (ii) Type-IV MI  & (ii)Type-III MI \\
   &&&&&&&&(iii) Type-V MI\\\\\hline\\
			Normal   & Yes  & Yes & Yes & Yes & Type-V MI& (i) Type-V MI  & 
			Type- V MI& (i) Type-II MI\\
				dispersion &   &  &  &    &  & (ii) Type-IV MI& & (ii) Type-III MI  \\
    dispersion &   &  &  &    &  &  & & (iii) Type-V MI  \\\\
			\hline
	\end{tabular}
 \smallskip
Convl: Conventional, Type-I MI: Primary((or) Symmetric) MI gain spectrum, Type-II MI: Asymmetric MI  sidebands, Type-III MI: Secondary MI spectrum, Type-IV MI: Peculiar MI gain spectrum and Type-V: Monotonically increasing gain. 
\end{table*}
\section{conclusion}
In conclusion, we have theoretically investigated the formation of MI gain spectra in a physical setting of fiber Bragg gratings with saturable nonlinearity and gain and loss. We have found that the obtained nonlinear dispersion curves do not exhibit the loop structure either in the upper branch or in the lower branch as opposed to the conventional systems due to the ratio assigned between the different nonlinearities. We have systematically classified our investigation of the MI gain spectrum based on the analysis of the dispersion curves into four cases, namely the bottom and top of the photonic bandgaps, and the anomalous and normal dispersions.
Having analyzed the dispersion relation first, we then examined the MI gain at the bottom of the photonic bandgap. The system has remarkably revealed the emergence of instability spectra with variations in the gain/loss parameter $g$ rather than the function of wavenumber, in the unbroken $\cal PT$ regime. It is worthwhile to mention that the finding of such a peculiar MI spectrum is new in the literature in the framework of any periodic structures. In addition, the CW manifests in a stable state at the exceptional point and asymmetric spectra tend to appear in the broken $\cal PT$-symmetric regime. On the other hand, the role of input power and saturable nonlinearity gives rise to the different MI spectra, namely the monotonically increasing gain which further rises with the increase in these parameters. It is interesting to note that in this unbroken $\cal PT$- symmetric regime the coupling coefficient has also exhibited the ramification of the peculiar MI bands. The peculiar spectrum persists in the top of the photonic band gap too accompanied by the conventional symmetric and asymmetric spectra on varying the value of gain/loss parameter. Though the coupling coefficient makes the system to experience the symmetric spectrum on either side of the zero wavenumber, the input power and saturable nonlinearity mainly disintegrate the symmetric ones into the multiple structures featuring a primary and secondary spectrum with a clear manifestation of discreteness in the sidebands. Note that in all the cases, peak gain rises as all the system parameters increase.

The system has shown diverse MI gain spectra for each $\cal PT$-symmetric regime when it comes to the anomalous dispersion regime. Although the gain/loss parameter has manifested in the complex spectrum when switching from the unbroken to broken $\cal PT$- symmetric regimes, the increase in both the gain/loss and coupling parameters substantially rise the peak gain of the sidebands in every regime, while the increase in the power and saturable nonlinearity suppresses the spectrum in the anti-Stokes wavenumber region. 
Conversely, in the normal dispersion regime, we have primarily observed the emergence of monotonically increasing gain in addition to the peculiar spectrum when we tune the value of both the coupling coefficient and the gain/loss parameter. In a similar way, the saturable nonlinearity and the input power cause the system to exhibit the monotonically growing gain which in turn translates into a conventional symmetric MI band when it operates in the above $\cal PT$ threshold.  It is important to stress that all of the instability spectra obtained in this study under various dispersion regimes and under $\cal PT$-symmetric conditions have qualitative differences from the ones obtained in a conventional Bragg grating structure. We hope that our findings open the door for future studies of localized modes such as Bragg solitons, using the synthetic grating structures that imprint the gain/loss profile.

\section*{Acknowledgement}
KT acknowledges the Department of Science and Technology (DST) and Science and Engineering Research Board (SERB), Government of India, through a National Postdoctoral Fellowship (Grant No. PDF/2021/000167). AG is supported by University Grants Commission (UGC), Government of India, through a Dr. D. S. Kothari Postdoctoral Fellowship (Grant No. F.4-2/2006 (BSR)/PH/19-20/0025). ML is supported by a DST-SERB through a National Science Chair (Grant No. NSC/2020/000029).

\end{document}